\documentclass[aps,pra,twocolumn,letterpaper,showpacs,superscriptaddress,floatfix,10pt]{revtex4-1}
\usepackage[pdftex]{graphicx}
\graphicspath{ {./images/} }
\usepackage{dcolumn}
\usepackage{bm}
\usepackage{bbm}
\usepackage[usenames, dvipsnames]{color}
\usepackage{comment}
\usepackage[colorlinks, linkcolor=blue]{hyperref}

\usepackage{layouts}

\usepackage[T1]{fontenc}

\usepackage{amsfonts}
\usepackage{amssymb}
\usepackage{amsmath}

\usepackage{amsthm}
\theoremstyle{remark}

\usepackage{multirow}

\usepackage[scr=boondoxo, scrscaled=1.05]{mathalfa}

\usepackage{overpic}

\usepackage{array}
\newcolumntype{L}[1]{>{\raggedright\let\newline\\\arraybackslash\hspace{0pt}}m{#1}}
\newcolumntype{C}[1]{>{\centering\let\newline\\\arraybackslash\hspace{0pt}}m{#1}}
\newcolumntype{R}[1]{>{\raggedleft\let\newline\\\arraybackslash\hspace{0pt}}m{#1}}
\usepackage{tabularx}

\def\avg#1{\mathinner{\langle{#1}\rangle}}

\newcommand\C{{\mathbb C}}
\newcommand\Z{{\mathbb Z}}

\newcommand\Lcal{\mathcal{L}}

\def\half{\tfrac{1}{2}}

\begin{document}

\title{Edge Theories for Anyon Condensation Phase Transitions}

\author{David M. Long}
\email{dmlong@umd.edu}
\affiliation{Department of Physics, Boston University, Boston, Massachusetts 02215, USA}
\affiliation{ARC Centre of Excellence for Engineered Quantum Systems\\School of Physics, The University of Sydney, Sydney, NSW 2006, Australia}
\affiliation{Condensed Matter Theory Center and Joint Quantum Institute,\\Department of Physics, University of Maryland, College Park, Maryland 20742, USA}

\author{Andrew C. Doherty}
\affiliation{ARC Centre of Excellence for Engineered Quantum Systems\\School of Physics, The University of Sydney, Sydney, NSW 2006, Australia}

\date{\today}

\begin{abstract}
The algebraic tools used to study topological phases of matter are not clearly suited to studying processes in which the bulk energy gap closes, such as phase transitions.
We describe an elementary \emph{two edge} thought experiment which reveals the effect of an anyon condensation phase transition on the robust edge properties of a sample, bypassing a limitation of the algebraic description.
In particular, the two edge construction allows some edge degrees of freedom to be tracked through the transition, despite the bulk gap closing.
The two edge model demonstrates that bulk anyon condensation induces symmetry breaking in the edge model.
Further, the construction recovers the expected result that the number of chiral current carrying modes at the edge cannot change through anyon condensation.
We illustrate the construction through detailed analysis of anyon condensation transitions in an achiral phase, the toric code, and in chiral phases, the Kitaev spin liquids.
\end{abstract}

\maketitle

\section{Introduction}
    \label{sec:intro}

    Motivated by experimental observation of the fractional quantum Hall effect~\cite{Tsui1982,Stormer1999,Wen2017}, and connections to fault tolerant quantum computing~\cite{Kitaev2003,Nayak2008,Lahtinen2017}, the study of \emph{topological phases of matter} has seen immense progress~\cite{Klitzing1980,Tsui1982,Thouless1982,Bednorz1986,Wen1990,Hasan2010,Wen2017,Berry1984,Kitaev2001,Kitaev2003,Kane2005a,Kane2005b,Kitaev2006,Kitaev2009,Kitaev2009a,Senthil2015,Ren2016,Kong2018,Kong2020,Kong2021,Kim2020}. These phases do not break any local symmetry---they have no local order parameter~\cite{Landau1937,Ginzburg1950,Landau1980}. Nonetheless, topological phases are sharply defined phases of matter. A tuning of parameters which takes a system from one topological phase to another must be accompanied by a divergence of the correlation length~\cite{Kitaev2009}---a conventional indication of a continuous phase transition~\cite{Landau1980}---or by a first order transition. Understanding topological phases required the development of new mathematical tools, and has driven many advances in our understanding of both condensed matter and other fields of physics~\cite{Wen2017,Klitzing2019,Braun1986,Jeckelmann2001}.

    \begin{figure}
        \centering
        \includegraphics[width=\linewidth]{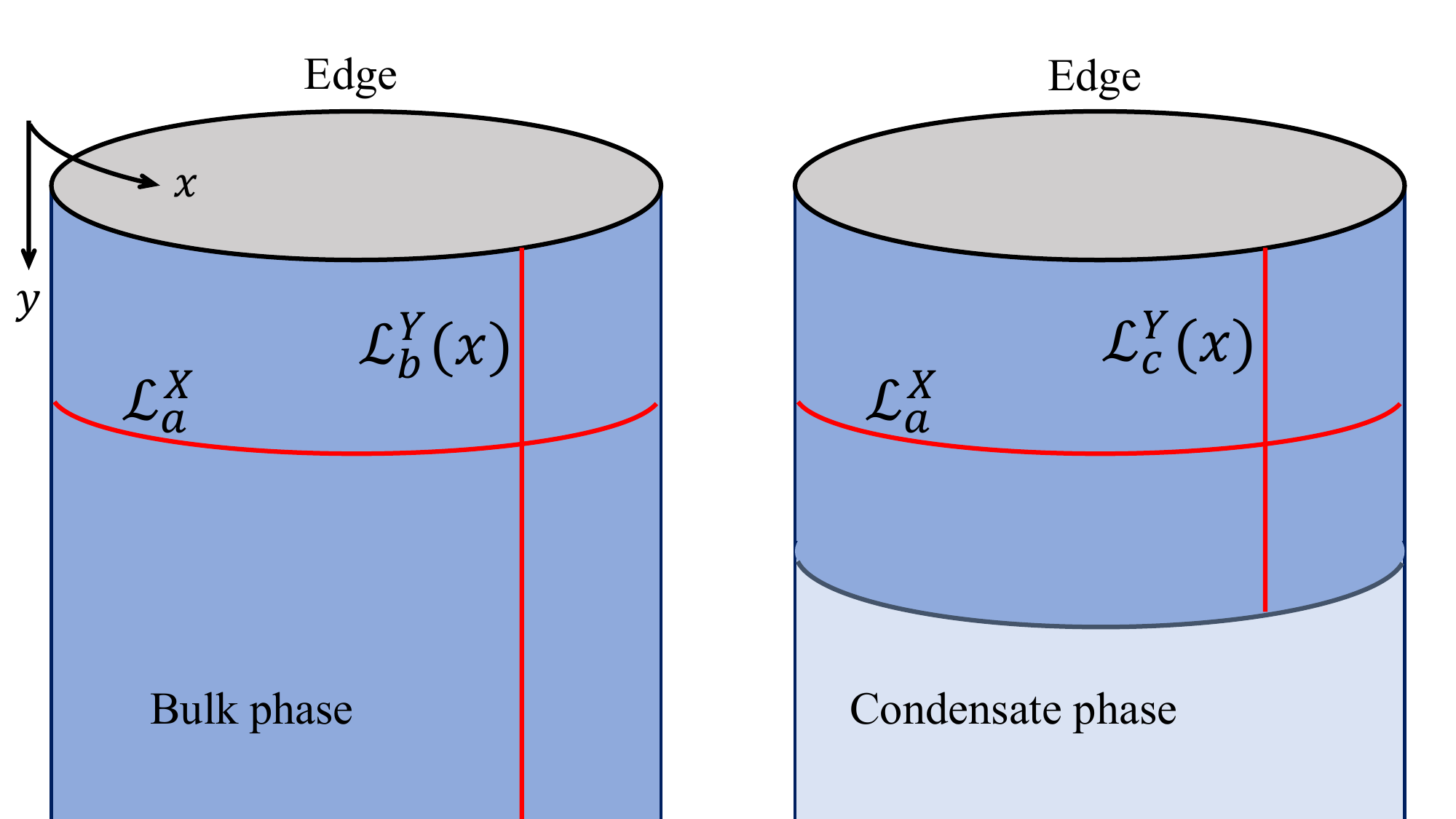}
        \caption{\label{fig:cond_edge}The existence of anyons in a semi-infinite cylinder of a topological phase implies the existence of nonlocal symmetry transformations on the edge~\cite{Lichtman2021}. The two edge model relates condensation of the anyon \(c\) to a breaking of this symmetry. Before condensation, the symmetry is implemented by unitary loop operators \(\Lcal_a^X\), corresponding to moving an anyon \(a\) around the circumference of the cylinder (the \(x\) direction). The vertical (\(y\) direction) anyon string operators, \(\Lcal_b^Y(x)\), also define nonlocal operations on the edge. After condensing an anyon \(c\), the edge in the condensate phase (right) is modeled by a thin strip of the uncondensed phase---forming two edges. The string operators tunneling \(c\) anyons between the two edges act locally on the new edge model, and explicitly break any \(\Lcal_a^X\) symmetry for which \(a\) and \(c\) have nontrivial braiding.}
    \end{figure}

    By now, there is a well developed mathematical toolkit for the study of gapped topological phases in two dimensions~\cite{Kitaev2006,Wen2017}. A particular focus of the literature is those phases that host \emph{anyons} as excitations in the bulk~\cite{Kitaev2003,Kitaev2006,Wen2017,Nakamura2020}. Another crucial piece of phenomenology is the gapless \emph{edge modes} that appear in finite samples of some topological materials~\cite{Thouless1982,Wen1991,Kane1997,Kitaev2006,Kong2018,Kong2020,Kong2021}. Important features of these propagating edge states are determined by the bulk topological phase, a feature known as the bulk-boundary correspondence.

    The study of phase transitions between topological phases has also seen progress, but is less well developed than the study of individual phases~\cite{Kosterlitz1973,Kosterlitz1974,Aharonov1984,Hansson2004,Burnell2018}. Restricting to an important class of phase transitions known as \emph{anyon condensation} transitions (also called topological symmetry breaking), far more is known~\cite{Bais2009,Barkeshli2010,Burnell2011,Burnell2012,Levin2013,Barkeshli2013,Eliens2014,Kong2014} (see Ref.~\cite{Burnell2018} for a review). In fact, there is a close analogy between anyon condensation and the conventional symmetry breaking transitions of the Landau-Ginzburg paradigm~\cite{Burnell2018}.

    Both the study of condensation phase transitions and of the bulk phase are usually framed in abstract algebraic terms~\cite{Kitaev2006,Kong2014}. While this abstraction makes it possible to make powerful conclusions about physical systems without reference to microscopic details, it is also a framework in which several natural physical questions are difficult or impossible to address.

    The motivating question of this work is: what happens to the edge modes when an anyon condenses in the bulk? An expert in topological phases might already intuit the result that the number of chiral edge modes remains the same~\cite{Kim2022modular}. However, this fact is not straightforward to see within the mathematical formalism of the anyon model~\cite{Kitaev2006,Kong2014} for the bulk, nor the conformal field theory (CFT) description of the edge~\cite{Ginsparg1988,Wen1994wave,Wen1994droplet,Francesco1997,Tong2016,Kong2018,Kong2021,Kong2020,Chatterjee2023}. Both descriptions rely on a restriction to just the low energy degrees of freedom in the model---they treat the bulk gap as being infinitely large. However, during a condensation phase transition, the bulk gap closes, and it becomes unclear this treatment is legitimate.

    We present an elementary and physically motivated construction that reveals the effect of anyon condensation on the edge of a topological phase. By performing a simple thought experiment in a \emph{two edge} geometry~\cite{Bais2009,Kitaev2011}---where the edge between the condensate phase and the vacuum is interrupted by a thin section of the uncondensed phase---we make a detailed characterisation of the edge of the condensate, and its relation to the phase before condensation. In particular, we show that bulk anyon condensation breaks a symmetry of a generic edge model, and that---when the edge theory is a CFT---anyon condensation in the bulk \emph{extends the chiral algebra} of the edge CFT~\cite{Francesco1997,Bais2009}. This process maintains the number of chiral modes in the CFT. Each of these results may be anticipated based on previous work~\cite{Bais2009,Kong2018,Kong2020,Kong2021}, but they are revealed straightforwardly with the two edge construction.

    The construction is illustrated in \autoref{fig:cond_edge}. In a cylinder geometry, the phase before condensation has nonlocal symmetry operators associated to loops of anyons~\cite{Nussinov2009sufficient,Gaiotto2015,Lichtman2021,McGreevy2023}. When an anyon \(c\) condenses, the ground state gains a macroscopic occupation of that anyon, and its associated loop operator can arise from the vacuum. Thus it acts trivially on the edge of the condensate model. Further, symmetries associated to anyons with nontrivial exchange statistics with \(c\) are explicitly broken by local string operators which tunnel \(c\) anyons between the two edges. Thus anyon condensation has a direct interpretation as symmetry breaking for the edge model~\cite{Burnell2018}. The same tunneling operators describe the extension of the chiral algebra when the edge theory is a CFT.

    Our construction explicitly addresses the fact that the bulk gap closes throughout the course of the transition. As the domain wall between the condensate phase and the uncondensed phase can always be gapped~\cite{Burnell2018}, low energy degrees of freedom can be identified unambiguously, so that control of the edge degrees of freedom can be maintained in the two edge model. The construction works without alteration both when the bulk anyons are Abelian or when they are non-Abelian, and for both chiral or achiral phases.

    Once the two edge geometry has been adopted, existing mathematical results can be used to characterize the new edge in great detail~\cite{Kong2018,Kong2020,Kong2021}. In particular, Ref.~\cite{Kong2018} explained how to calculate the new algebraic data describing the edge of the condensate phase~\cite[Eq.~(5.3)]{Kong2018}\footnote{It is not clear to us what specific manipulation of \cite[Eq.~(5.3)]{Kong2018} is needed to reproduce our results, but we believe such a calculation should be possible.}, and how to compute long-wavelength observables on the new edge. Reference~\cite{Chatterjee2023} also formalises the notion in which anyon condensation corresponds to symmetry breaking on the edge. Our heuristic construction reproduces parts of these results, but is substantially more elementary.

    Not all features of the edge are uniquely determined by the bulk topological phase. As such, our analysis should be interpreted as restricting the possible edge theories that may be realised by the condensate, given a particular set of edge properties before the condensation. In principle, additional structure not captured by our analysis may occur in the edge theory as a result of fine tuning of the edge degrees of freedom. We will restrict ourselves to the low energy features of the edge that are robust to generic perturbations, which may be characterised in some detail.

    This paper is structured as follows. In \autoref{sec:background}, we provide an intuitive overview of basic notions in the study of topological phases of matter. Our discussion avoids most of the abstract mathematical machinery of this field, and instead emphasises the qualitative features required to follow the two edge thought experiment, which is presented in \autoref{sec:general}. In Secs.~\ref{sec:toric} and \ref{sec:KSL}, we make the construction more concrete by making a detailed exploration of anyon condensation in two examples: the toric code and the Kitaev spin liquids, respectively. We discuss the implications of our results in \autoref{sec:disc}.

\section{Background}
    \label{sec:background}

    Understanding the different possible phases of matter, and the phase transitions between them, is one of the overarching goals of condensed matter physics~\cite{Landau1980,Wen2017}. Many phases can be successfully characterised by the different symmetries they manifest~\cite{Landau1937,Ginzburg1950}. Phase transitions are then understood as being due to the (spontaneous) breaking of such symmetries, as revealed by a local order parameter attaining a nonzero expectation value. This is known as the Landau-Ginzburg paradigm~\cite{Landau1937,Ginzburg1950}.

    In recent decades, it has been appreciated that not all phases (nor phase transitions) can be understood within the Landau-Ginzburg framework~\cite{Wen2017}. \emph{Topological phases of matter} (\autoref{sec:topo_phases}) cannot be distinguished from nontopological phases by any measurement of a local observable---they have no local order parameter. (They do have nonlocal symmetries~\cite{McGreevy2023,Lichtman2021}.)

    Phase transitions between different topological phases are, in general, complicated~\cite{Burnell2018}. The primary topic of this paper will be anyon condensation phase transitions (\autoref{sec:anyon_cond}), a particularly simple class of phase transitions with many analogies to Landau-Ginzburg symmetry breaking transitions. Indeed, the constructions of \autoref{sec:general} will make these analogies quite precise when we focus on the effect of condensation on the edge modes.

    \subsection{Topological phases of matter}
        \label{sec:topo_phases}

        What different authors mean by a topological phase of matter sometimes depends on context~\cite{Wen2017}. Here, we will mean a zero temperature phase of a two dimensional gapped bosonic system (for instance, a lattice of spins) which differs from the trivial vacuum phase (containing uncoupled product states) in nonlocal degrees of freedom. We will assume that the reader is familiar with basic notions regarding anyons, including fusion and braiding~\cite{Kitaev2006,Wen2017}.

        Beyond introducing notation, our focus will be on the nature of domain walls between topological phases---especially domain walls to the vacuum phase, which we call edges~\cite{Kitaev2012}. The bulk anyon theory~\cite[Appendix~E]{Kitaev2006} and the nature of the edge actually completely characterise all topological phases~\cite{Lan2016}.

        \subsubsection{Anyons}
            \label{subsec:anyons}

            The quasiparticle excitations of a topological phase, called \emph{anyons}, play a key role in their phenomenology~\cite{Wen2017}. For instance, they are crucial in the description of the fractional quantum Hall effect~\cite{Stormer1999}. They have also been proposed as a tool to design fault tolerant quantum computers~\cite{Kitaev2003}. This section briefly summarises the properties of anyons that will be needed to understand the rest of the paper---fusion and braiding. The reader may consult Ref.~\cite[Appendix E]{Kitaev2006} for a complete description of the algebraic theory of anyons, called a \emph{unitary braided fusion category} (UBFC).

                \emph{Fusion.}---An anyon theory is based on a finite set of anyon species, \(\{1, a, b, \ldots, c\}\). These species should be viewed as superselection sectors: they label the nonlocal properties of an excitation. As such, they are sometimes referred to as \emph{topological charges}.

                \begin{figure}
                    \centering
                    \begin{overpic}[width=\linewidth]{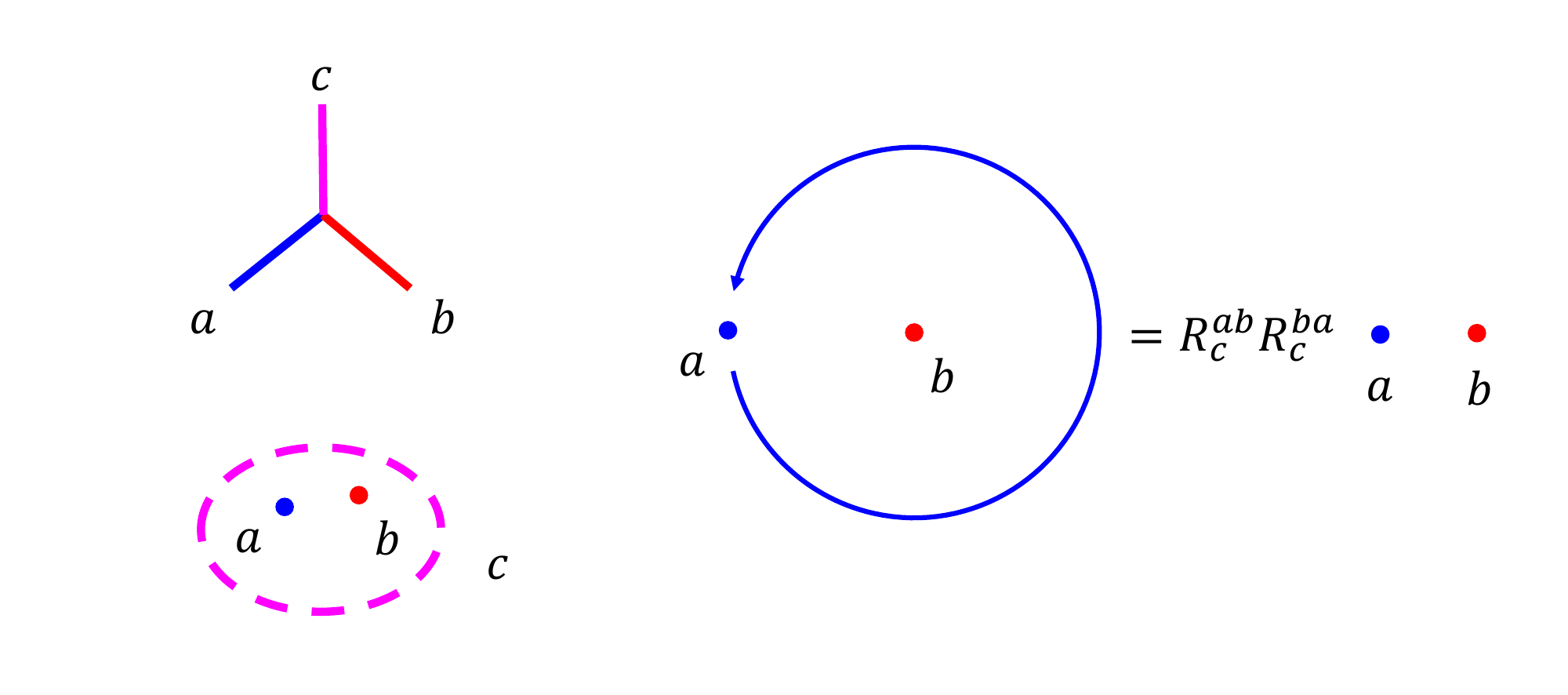}
                    \put(5,40){(a)}
                    \put(5,10){(b)}
                    \put(40,30){(c)}
                \end{overpic}
                    \caption{\label{fig:fuse_braid}Fusion of anyons, \(a\times b = c\), has two interpretations: (\textbf{a}) two anyons collide and form a new \(c\) bound state, or (\textbf{b}) a region of the system contains two anyons, such that the total topological charge of the region is \(c\). (\textbf{c}) Braiding an anyon \(a\) around \(b\) introduces a phase to the wave function, \(e^{i \theta} = R^{ab}_c R^{ba}_c\). If \(a\) and \(b\) are non-Abelian, the phase may depend on the fusion channel, \(c\).}
                \end{figure}

                Fusion of anyons has two interpretations. Either two anyons coalesce to form another anyon, or alternatively two anyons (of topological charge red and blue, say) placed in proximity appear like a different anyon (purple) from far away (\autoref{fig:fuse_braid}). The mathematical description of fusion is the same for both physical pictures.

                The notation
                \begin{equation}
                    a \times b = c
                \end{equation}
                is used to express that anyon species \(a\) and \(b\) fuse to an anyon \(c\). It may happen that the result of fusing \(a\) and \(b\) results in one of several possible anyons, depending on the state of the rest of the system. Such anyon theories are called non-Abelian, and their fusion rules are written
                \begin{equation}
                    a \times b = \sum_c N_c^{ab} c,
                    \label{eqn:nonab_fuse}
                \end{equation}
                where \(N_c^{ab}\) are integers giving the multiplicity of channels in which \(a\) and \(b\) fuse to \(c\). Thus the sum here should be interpreted as a direct sum. In fact, there is a close analogy between this algebraic structure and the decomposition of tensor products of group representations into a direct sum of irreducible representations~\cite[Appendix~E]{Kitaev2006}.

                Among the anyons is a distinguished anyon \(1\), the vacuum anyon. This is a formalization of the absence of an anyon. It fuses trivially with all other anyons:
                \begin{equation}
                    1 \times a = a \times 1 = a.
                \end{equation}

                \emph{Braiding.}---The exchange of identical particles introduces a phase factor to the wave function which, in two dimensions, need not be real but can be any phase factor (hence the name anyon). The same is true of any other permutation of anyons that leaves the final state indistinguishable from the original. For instance, moving an anyon \(a\) in a complete circle around another anyon \(b\).

                The rules for how the wave function changes are encoded in a set of coefficients \(R^{ab}_c\). These are unit modulus complex numbers that assign a phase to the exchange of \(a\) and \(b\) in the background state where they fuse to \(c\). If \(a \neq b\), the resulting state is distinguishable from the initial state, so the individual \(R^{ab}_c\) do not have a gauge invariant meaning. Several topological invariants may be computed from the \(R\)s, but we will focus on the \emph{monodromy}, and the \emph{topological spin}.

                The monodromy associated to \(a\) and \(b\) in fusion channel \(c\) is the phase \(R^{ab}_c R^{ba}_c\), which measures the phase due to wrapping \(a\) in a circle around \(b\) (\autoref{fig:fuse_braid}).

                The topological spin is a feature of a single anyon \(a\). It is also a phase factor, denoted \(\theta_a\). It should be thought of as analogous to the usual spin---encoding the phase acquired by a particle when it is rotated~\footnote{The precise definition is slightly more involved, as this phase may depend on the state of the rest of the system, and \(a\) need not have an actual spin associated to rotations.}. Bosons all have \(\theta_a = 1\), while fermions have \(\theta_a = -1\). In general, \(\theta_a\) may be any rational phase.

        \subsubsection{Edge modes}
            \label{subsec:edge_modes}

            In the bulk of a topological phase, there are (by definition) no excitations within the energy gap. However, at domain walls between different topological phases, there may exist dispersing modes that traverse the gap, and may carry a current~\cite{Klitzing1980,Thouless1982,Hasan2010,Wen2017,Kong2018,Kong2020,Kong2021}. This feature has been recognised for much longer than the existence of anyons, going all the way back to the integer quantum Hall effect (in the fermionic context)~\cite{Klitzing1980,Thouless1982}. Many important features of the edge modes are determined by the bulk topological phase, a relation which is known as the bulk-boundary correspondence~\cite{Kong2021}.

            For the edge modes to scatter off of some impurity at the edge, there must be another state for them to scatter \emph{to}. As the edge modes are in the middle of a gap in the bulk, there are no other states with which they can hybridise. It is said that the edge modes are protected by the bulk gap. If the bulk gap closes then the edge modes may, in principle, meander into the bulk. This is why the edge modes can change through a bulk phase transition, where the gap closes.

            When the domain wall between the phases has gapless modes, the low energy effective theory for these excitations can often be described by a \emph{conformal field theory} (CFT)~\cite{Ginsparg1988,Francesco1997,Tong2016}, and we will restrict our analysis of gapless edges to the cases where this is possible. CFT is a huge subject on its own, and the modern theory of anyons is in large part a descendant of this field~\cite{Moore1988,Moore1989,Moore1990}. We will be able to avoid the vast majority of CFT machinery, but we will need to understand the connection between a bulk anyon theory and the CFT at its edge. The interested reader should consult Refs.~\cite{Ginsparg1988,Francesco1997} for more technical and complete information on CFTs.

            Any CFT has several quantities associated to it, including: its \emph{central charge}, a list of \emph{primary fields}, and a \emph{chiral algebra} of operators that can be implemented locally, and correspond to observables~\cite{Fuchs1997}. We will present an intuitive explanation of these quantities, and their relation to the bulk anyon model.

            Much of this data is determined by the bulk anyon theory (the UBFC), but the central charge is not. Heuristically, a CFT may be divided up into a number \(c\) of right-moving bosonic modes and \(\bar{c}\) left-movers. The central charge need not be an integer when the modes do not have a bosonic character. The combination \(c + \bar{c}\) appears in formulae for the heat capacity of the model, while the \emph{chiral central charge} \(c_- = c - \bar{c}\) measures how many more right-movers there are than left-movers. A nonzero \(c_-\) implies the theory has an inherent handedness, and breaks time-reversal symmetry. The chiral central charge manifests physically in chiral heat currents at low temperatures~\cite{Read2000,Kitaev2006}.

            The bulk anyon data turns out to only constrain \(c_- \mod 8\)~\cite{Kitaev2006}. There are distinct topological phases, with different \(c_-\) at their edges that nonetheless have the same anyon content. A complete classification of bosonic topological phases requires both a description of the anyons, and the value of \(c_-\)~\cite{Lan2016}.

            The other CFT data relevant to our discussion have analogies in the anyon theory. First, we consider the chiral algebra~\cite{Ginsparg1988,Fuchs1997,Bais2009}. In a colloquial description, the chiral algebra is the collection of local operators~\footnote{This characterization can be obtained from more formal definitions by noting that the chiral algebra generators are constructed from commutators (or operator product expansions) between the stress tensor (which is a local operator) and local symmetry generators. For the simplest example of the Virasoro algebra, the construction of the symmetry generators (usually denoted \(L_n\)) from the stress tensor is standard~\cite{Ginsparg1988}. For more complicated chiral algebras, such as Kac-Moody algebras, similar constructions exist~\cite{Ginsparg1988,Fuchs1997}.}. The theory of anyons ignores all local operations, so the subsequent conclusions we draw about the relation between anyons and CFT should all be considered modulo the chiral algebra. Equipping ourselves with some notation, we will write \([\phi]\) for the equivalence class of the edge field \(\phi\) under the chiral algebra.

            CFTs in two dimensions have a very large symmetry algebra, which allows many exact analytic results~\cite{Ginsparg1988,Francesco1997}. The primary fields of a CFT are the most relevant (in the renormalisation group sense) fields in each symmetry block. When the only symmetries considered are those spatial symmetries common to all CFTs, the primaries are called Virasoro primaries. Less relevant operators in the same symmetry block are called descendants of the primary field.

            The primary field equivalence classes \([\phi]\) correspond to the anyons of the bulk theory~\cite{Tong2016}. Indeed, moving an anyon from the bulk onto the edge implements a primary field operator in the edge CFT, up to local details. This construction is the origin of most relations between CFT and anyons, and will be important for characterising the effects of anyon condensation.

            An important parameter assigned to each primary field are its \emph{conformal weights}, \(h_{\phi}\) and \(\bar{h}_\phi\). These are the eigenvalues of the right- and left-moving parts of the primary field \(\phi\) under rescaling of space, which is an important symmetry of CFTs. The conformal weights are closely related to the topological spin \(\theta_\phi\) of the bulk anyon corresponding to \(\phi\). We have
            \begin{equation}
                \theta_{\phi} = e^{2\pi i (h_\phi-\bar{h}_\phi)}.
            \end{equation}

            \begin{figure}
                \centering
                \includegraphics[width=\linewidth]{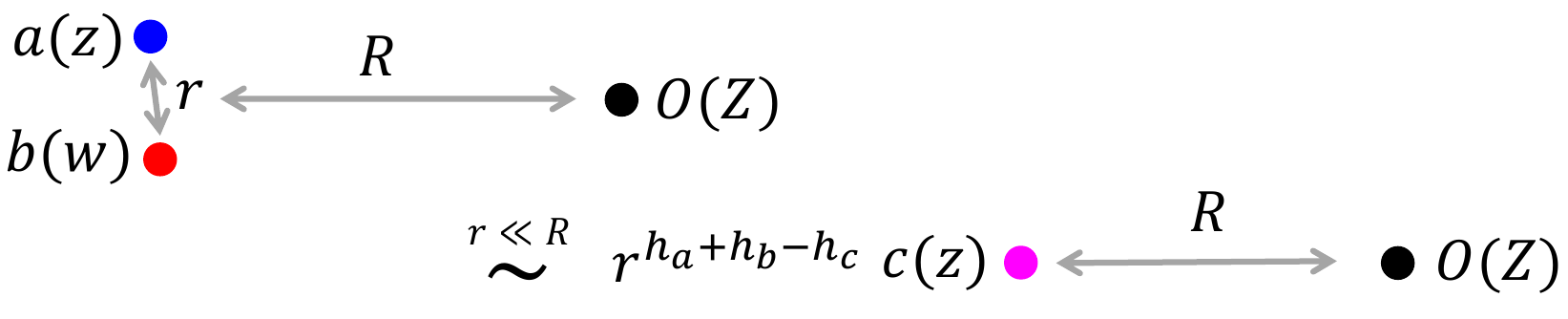}
                \caption{\label{fig:OPE}When two primary fields \(a(z)\) and \(b(w)\) are very close, they present as a single field \(c(z)\) to a more distant \(O(Z)\) for the purpose of correlation functions. This is the basis for the operator product expansion, Eq.~\eqref{eqn:OPE}. Note the similarity to the picture of anyon fusion in \autoref{fig:fuse_braid}.}
            \end{figure}

            The fusion and braiding of pairs of anyons appears in the CFT language within the \emph{operator product expansion} of primary fields. We will consider just the holomorphic component (associated to right-movers) of the CFT. Suppose one measures a correlator between a pair of nearby primary fields, \(a(z)\) and \(b(w)\) (with two dimensional coordinates parameterised by the complex numbers \(z\) and \(w\)), with a more distant field, \(O(Z))\). Taking \(z \to w\), we can treat \(a(z) b(w)\) as composing a single composite field. Writing \(c(z)\) for the most relevant primary making up this field, we have (\autoref{fig:OPE})
            \begin{equation}
                \avg{a(z)b(w)O(Z)} \sim (z-w)^{h_a + h_b - h_c} \avg{c(z) O(Z)}.
                \label{eqn:OPE_full}
            \end{equation}
            Here, the factor of \((z-w)^{h_a + h_b - h_c}\) is introduced to ensure both sides of the equation have the same behaviour under a rescaling of space. The numbers \(h_{a,b,c}\) are the conformal weights.

            We use a common shorthand for Eq.~\eqref{eqn:OPE_full},
            \begin{equation}
                a(z)b(w) \sim (z-w)^{h_a + h_b - h_c} c(z), 
                \quad\text{or}\quad
                [a][b] = [c].
                \label{eqn:OPE}
            \end{equation}
            This is the analogous expression to \(a \times b = c\) for anyons. Indeed, the physical picture associated with the OPE---putting two primary fields close together and measuring their behaviour from far away---is reminiscent of anyon fusion.

            (More than one primary field may appear in the OPE of \([a]\) and \([b]\), as was the case for anyons. We will not discuss any such models.)

            Braiding also follows from \eqref{eqn:OPE}. Taking \(a(z)\) in a loop around \(b(w)\) returns the right hand side to itself up to a phase associated to possibly changing branches in the Riemann surface of \((z-w)^{h_a + h_b - h_c}\). This phase factor is \(R^{ab}_c R^{ba}_c\).

    \subsection{Anyon condensation}
        \label{sec:anyon_cond}

        The Landau-Ginzburg paradigm associates continuous phase transitions to the spontaneous breaking of symmetry. In general, transitions between topological phases do not fit within this framework. However, an important subclass of these transitions, called anyon condensations, do have some analogy with symmetry breaking. This analogy has prompted some authors to call anyon condensation transitions ``topological symmetry breaking''~\cite{Burnell2018}.

        The quotation marks are usually mandatory. Topological features are not associated to any local symmetry of the model, and there is no local order parameter which acquires a nonzero expectation value in the ``topological symmetry broken'' phase. (There may be a nonlocal one.) Even so, the study of anyon condensation is a simple starting point for any theory of topological phase transitions.

        \subsubsection{Bulk description}
            \label{subsec:anyon_bulk}

            The consequences of anyon condensation are simple to explain in the bulk. Their effect on the edge of a system will be deduced in \autoref{sec:general}. For a more detailed review than presented here, the interested reader should consult Ref.~\cite{Burnell2018}.

            Glibly, anyon condensation is Bose condensation of anyons. Through some tuning of potentials, the ground state acquires a macroscopic population of anyons of species \(c\). For this to occur, there cannot be a Pauli exclusion associated to \(c\). That is, \(c\) must be symmetric under exchange---it is a self-boson. In the condensate, the creation of \(c\) anyons from the vacuum is free---it is a local operation with no energy cost.

            The condensed anyon \(c\) need not be a \emph{mutual} boson with other anyons in the theory. There may be nontrivial statistics between \(c\) and some other anyon species, \(b\). However, if this does occur, then movement operators for \(b\) must not commute with \(c\) creation operators. As \(c\) creation is now a local operation, the background condensate of \(c\) anyons in the ground state means that it now costs energy to move a \(b\) anyon around the system. Indeed, the cost of creating an anyon-antianyon pair \(b \bar{b}\) and moving \(b\) along some trajectory away from \(\bar{b}\) scales linearly in the length of the trajectory. This provides a linear confining potential between the anyon-antianyon pair, just as occurs between pairs of quarks in quantum chromodynamics. Thus the \(b\) anyon never appears in isolation, only close to its antianyon \(\bar{b}\) near which it remains \emph{confined}.

            The confined anyon \(b\) should not be regarded as being a part of the condensate anyon theory. The anyon theory concerns only nonlocal features, and \(b\) always comes with its antiparticle pair, and so appears trivial from far away.

            Further consequences arise from the background of \(c\) anyons. Anyons \(a\) and \(a' = c \times a\) related by the fusion of \(c\) become identified in the condensate model. As the creation of \(c\) anyons is free, a local operation now relates \(a\) and \(a'\). The anyon theory only keeps track of nonlocal features, so \(a\) and \(a'\) must be regarded as the same anyon after condensation.

            The inverse process of the identification of anyons may also occur upon condensation. An anyon \(d\) may decompose into two (or more) anyons \(d_i\). This occurs when \(d\) and its antiparticle \(\bar{d}\) can fuse to \(c\): \(N^{d \bar{d}}_c > 0\). The anyon \(c\) should now be identified with the vacuum, so \(d\) and \(\bar{d}\) may annihilate in two distinct ways. This turns out to violate the consistency conditions for an anyon theory. The resolution is that \(d\) dissociates in the new phase. Each dissociated anyon \(d_i\) has only one channel fusing to the vacuum with its antianyon \(\bar{d}_i\).

            These rules---confinement, identification, and splitting---will be enough for us to work with. Anyon condensation also has a description in terms of the mathematical machinery of a UBFC, but we will avoid this formalism in favour of a more physically-based picture.

        \subsubsection{Gapped domain walls}
            \label{subsec:gapped_dw}

            There is a close connection between anyon condensation and gapped domain walls between topological phases~\cite{Kitaev2012,Kong2014}. Indeed, two topological phases are related by anyon condensation if, and only if, the domain walls between them can be gapped~\cite{Kitaev2012,Levin2013,Kong2014,Burnell2018}. (More precisely, each phase can be obtained by condensation from some parent phase.) This crucial fact underlies our analysis of the edge of a topological phase as it goes through a condensation phase transition. It also allows for a description of how anyons behave when moving through a domain wall in terms of the confinement, identification and splitting rules from \autoref{subsec:anyon_bulk}.

                \emph{Phases related by condensation.}---The domain walls between two topological phases can be gapped if, and only if, the two phases are related by anyon condensation~\cite{Kitaev2012,Levin2013,Kong2014,Burnell2018}. We will not give the full proof of this statement, but it is revealing that the proof relies not on UBFC technology, but rather on CFT. The UBFC, while it strongly constrains the edge, is not sufficient to completely characterise it. This is also the case when considering the effect of anyon condensation on the edge---the UBFC description of condensation leaves the resulting edge modes ambiguous.

\begin{figure}
                    \centering
                    \includegraphics[width=\linewidth]{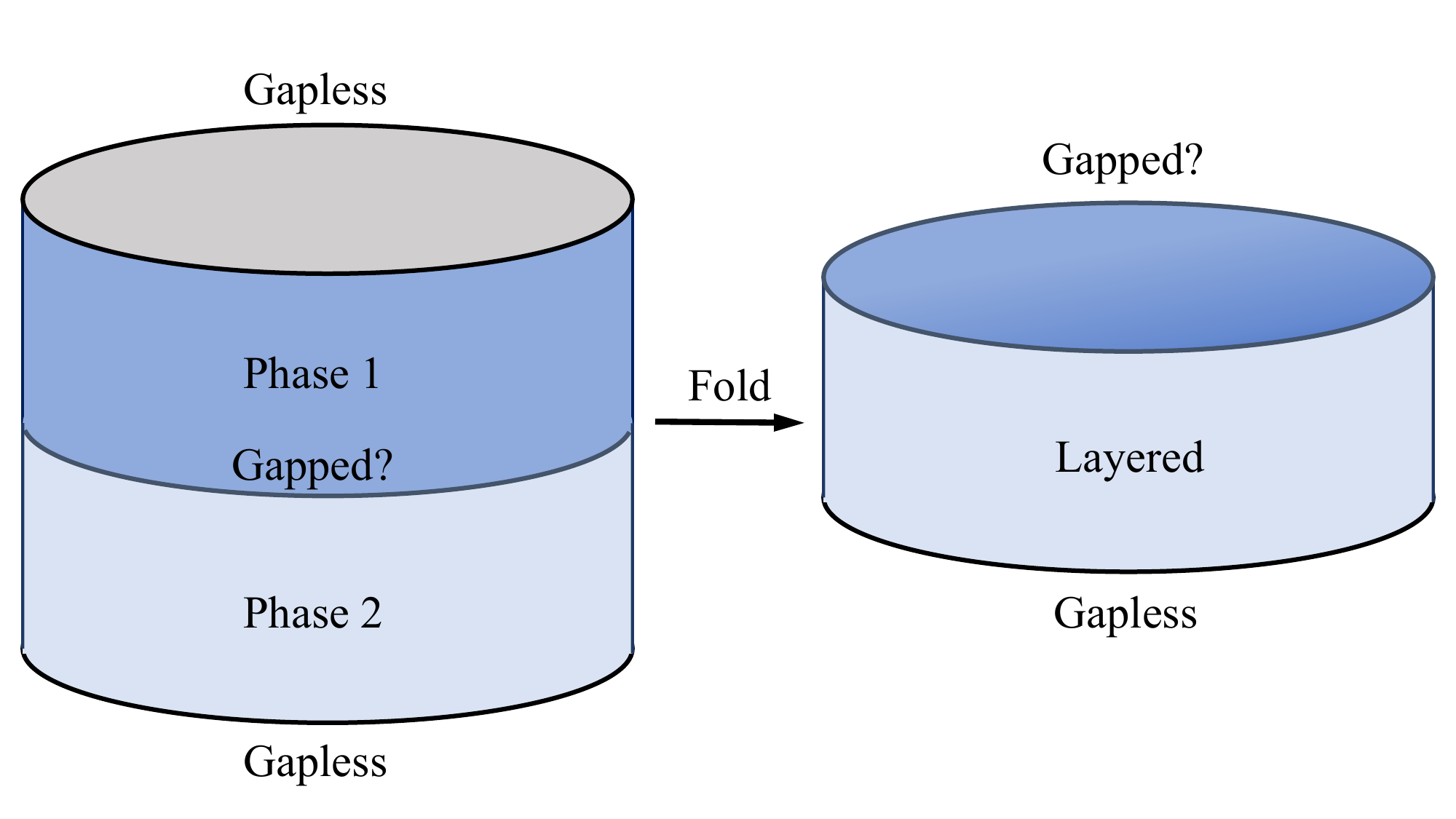}
                    \caption{\label{fig:gapped_folding}A domain wall between two topological phases (1 and 2) can be gapped if, and only if, the two phases are related by an anyon condensation transition. In a cylindrical geometry, the presence of a gapped domain wall is equivalent to the presence of a gapped edge after folding the cylinder into two layers. The gapless bottom edge must be described by a CFT that can be achieved in a one dimensional Hamiltonian system, which puts constraints on the bulk through the bulk-boundary correspondence. These constraints turn out to be that the phases 1 and 2 are related by anyon condensation.}
                \end{figure}

                At the level of a sketch, the proof that the domain walls may be gapped proceeds as follows. Consider a domain wall between two topological phases in a finite cylindrical geometry, such that their edges to vacuum are gapless (which may or may not require fine tuning). By folding the cylinder, as illustrated in \autoref{fig:gapped_folding}, only two edges need to be considered: a gapless bottom edge, and a top edge that may or may not be gapped. The cylinder now consists of a double layer of topological materials.

                When can the top edge have a gap? If the top edge can be gapped, then the bottom edge is the only gapless thing left in the system. Shrinking the cylinder produces a quasi-one-dimensional gapless model, described by some CFT. There are strong constraints on what CFTs can actually occur in one-dimensional Hamiltonian models, so the top edge can only be gapped if the bottom edge meets these constraints.

                However, the CFT at the edge of a topological phase is (barring fine tuning) determined by the topological order in the bulk (\autoref{subsec:edge_modes})---there is a bulk-boundary correspondence. A constraint on the edge CFT can then be extended to a constraint on the double-layered topological phase. These constraints turn out to require that the two phases in the unfolded cylinder were related by anyon condensation. More precisely, two gapped topological phases \(\mathcal{A}\) and \(\mathcal{B}\) may share a gapped edge if there exists a third phase, \(\mathcal{C}\), such that both \(\mathcal{A}\) and \(\mathcal{B}\) can arise as condensates of \(\mathcal{C}\). That is, there is some set of anyons that can be condensed in \(\mathcal{C}\) which drives a transition into the phase \(\mathcal{A}\), and there is another (distinct, if \(\mathcal{A} \neq \mathcal{B}\)) set of anyons which when condensed cause a transition from \(\mathcal{C}\) to the phase \(\mathcal{B}\)~\cite{Kong2014,Burnell2018}.

                The reverse implication is also true: if the two phases are related by anyon condensation, then the domain wall between them can be gapped.

                We will explore the consequences of these gapped domain walls in \autoref{sec:general}.

                \emph{Anyons moving through domain walls.}---The effect of moving through a gapped domain wall on an anyon can be understood through the rules that appeared in the algebraic description of condensation: confinement, identification and splitting~\cite{Burnell2018}. We will focus just on confinement and identification.

                If an anyon in the uncondensed phase is confined in the condensate, then it can not move freely through the domain wall. Instead, it becomes trapped at the boundary, leaving behind an excitation. In this context, we also describe it as being confined.

                Alternatively, if the anyon in question is a mutual boson with the condensed particle \(c\), it is deconfined in the condensate. The background of \(c\) anyons in the condensate now means that the anyon can fuse with any number of \(c\)s, and all such fusion products become identified. Any two anyons which are identified in the condensate become indistinguishable when they cross the domain wall.

                A particularly striking example of this is moving \(c\) itself through the domain wall. The anyon \(c\) becomes identified with the vacuum---in a more physical picture, it is absorbed into the condensate background.

                The reverse process can also happen. Anyons \(c\) can emerge from the condensate to the uncondensed phase, without needing a \(\bar{c}\) antianyon to be present in the uncondensed domain. The antianyon remains in the condensate background.

                When the gapped domain wall is to the vacuum phase---when it is an edge---all anyons either become condensed or confined at the edge. This is a simple consequence of the fact that the vacuum phase only supports the vacuum anyon.

\section{Two edge construction for condensation edge effects}
    \label{sec:general}

    In this section we present the abstract characterization of bulk anyon condensation's effect on edge physics. Our understanding is based around a model for the edge that relates the presence of bulk anyons to symmetries of the low energy edge model~\cite{Lichtman2021}. In a \emph{two edge} geometry (\autoref{fig:cond_edge}), the model can be used to characterise the effects of bulk anyon condensation on the edge.

    We begin by describing the two edge model in \autoref{sec:two_edge}. This model is a generalization of a model in Ref.~\cite{Lichtman2021}, adapted to be appropriate for the study of bulk phase transitions~\cite{Bais2009,Kitaev2011}. Interrogation of this model provides elementary methods to find general consequences for the edge in terms of the properties of bulk anyons. Several such consequences are found in \autoref{sec:consequences}.

    \subsection{Two edge model}
        \label{sec:two_edge}

        The two edge model is based around the analysis of a particular geometry of a topological phase of matter---a semi-infinite cylinder~(\autoref{fig:cond_edge}). This geometry is convenient because of its simple edge geometry, a single circle, and because the non-contractible loops wrapping around the circumference of the cylinder can be used to access topological information, provided the circumference is much larger than the correlation length.

        The eponymous two edges will refer to a domain wall between a condensate phase and an intermediary uncondensed phase, and a second edge between the uncondensed phase and the vacuum, but a nontrivial analysis of edge physics is possible even with only the single edge shown in the left of \autoref{fig:cond_edge}. This analysis was pursued in the context of achiral phases of matter in Ref.~\cite{Lichtman2021}, and in \autoref{subsec:one_edge}, we reiterate it in a form adapted to also apply to chiral phases of matter, before considering condensation transitions in \autoref{subsec:2nd_edge}.

        \subsubsection{One edge}
            \label{subsec:one_edge}

            The central observation of the analysis of Ref.~\cite{Lichtman2021} is that the existence of bulk anyons implies the presence of symmetries in any effective model of the edge. In this section, we make a minor extension to that argument which allows us to address chiral topological phases.

            We first demand two natural requirements of any effective edge Hamiltonian. The first is locality: the edge Hamiltonian must be a sum of terms that act only within a finite range (in the continuum they only act at a point). The second is that the effective edge Hamiltonian, \(H_{\mathrm{edge}}\), does not excite the bulk gap. That is, the edge model we consider is capturing only the low energy degrees of freedom of the edge.

            With these assumptions, nontrivial conclusions about the nature of the edge model are possible. Consider creating an anyon \(a\) and its antianyon pair \(\bar{a}\) from the vacuum at a distance \(y\) from the edge, with \(y \gg \xi\) the correlation length. Transport \(a\) around the cylinder (in the \(+x\) direction, say) and then annihilate the \(a\bar{a}\) pair with each other. Call this operation \(\Lcal^X_a\).

            The string operator \(\Lcal^X_a\) does not change the number of excitations in the bulk: it returns the bulk to the vacuum state after annihilating the excitations \(a\) and \(\bar{a}\) with each other. As such, \(\Lcal^X_a\) acts as a unitary on the low energy degrees of freedom of the material. All these degrees of freedom are at the edge, as the bulk is gapped. Due to the strongly-correlated nature of the phase, this action may be nontrivial on the edge degrees of freedom, even if the entire pair creation and annihilation process took place far away from the edge. However, due to the finite correlation length in the material, and the assumption of locality of the edge Hamiltonian, we still have that the norm of the commutator \(\|[\Lcal^X_a,H_{\mathrm{edge}}]\|\) decreases exponentially in \(y/\xi\). That is, \(\Lcal^X_a\) acts as a symmetry transformation on \(H_{\mathrm{edge}}\) if \(y \gg \xi\).

            Additional details of the symmetry action can be worked out by considering the string operator \(\Lcal_b^Y(x)\). This operator takes \(b\) from very far away in the bulk, along the vertical line at position \(x\), and off the edge.

            To describe the action of \(\Lcal_b^Y(x)\) on the effective edge model, we consider three different cases. First, we consider the case of the edge being gapped (so the bulk phase is achiral). Recall from \autoref{sec:background} that all the bulk anyons are either confined or condensed at a gapped edge. The trivial case is that of \(b\) being confined at the edge. Then, \(\Lcal_b^Y(x)\) does not act just on the low energy degrees of freedom: it leaves behind an excitation, which can easily be on the same scale as the bulk gap.

            \begin{figure}
                \centering
                \includegraphics[width=\linewidth]{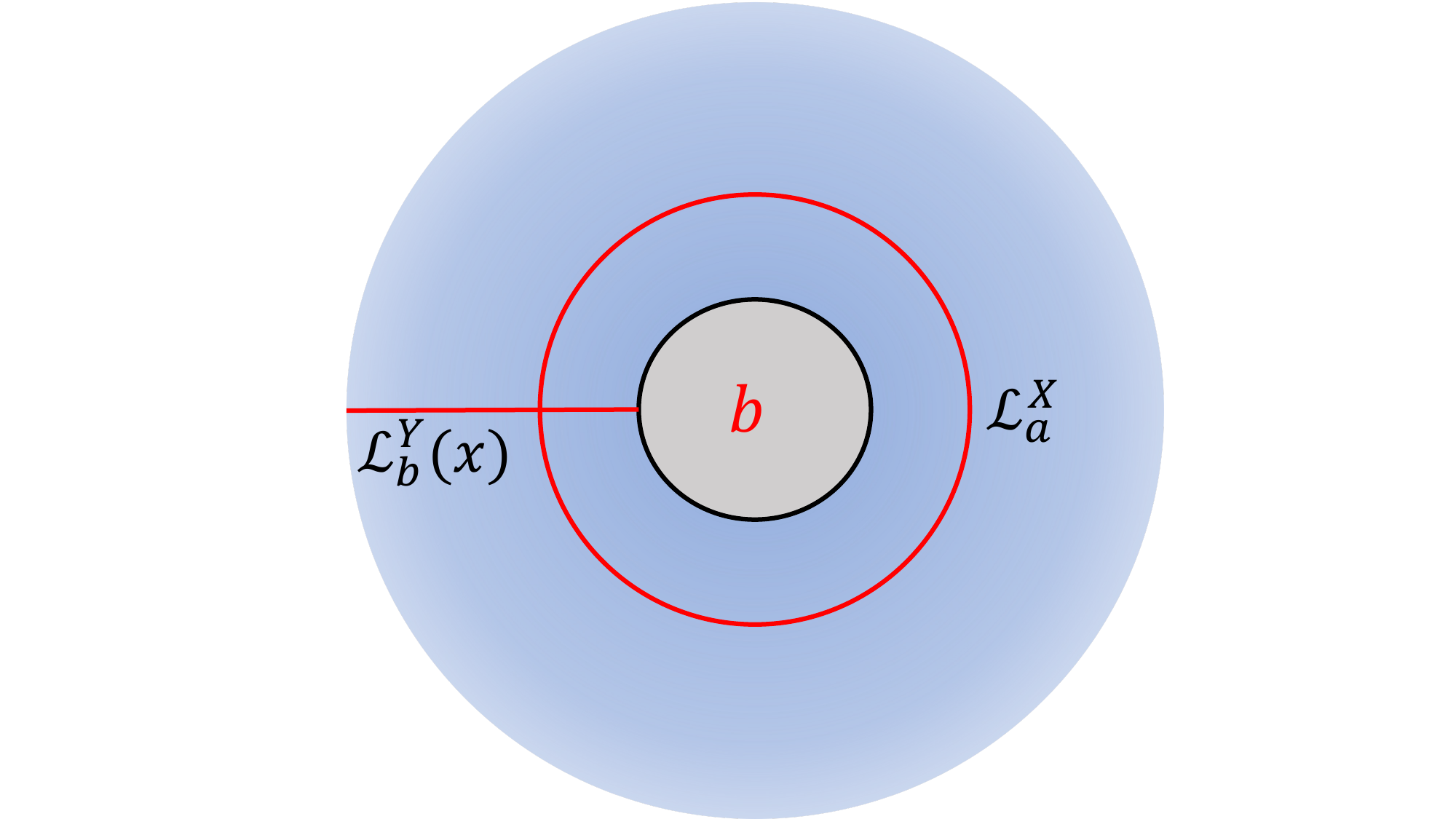}
                \caption{\label{fig:punctured_plane}A deformation of the cylindrical geometry shown in \autoref{fig:cond_edge} is a flat plane with a hole cut out.}
            \end{figure}

            On the other hand, when \(b\) is condensed at the edge, \(\Lcal_b^Y(x)\) does not leave behind any bulk excitation, and thus has an effective action on the edge. The nature of this action is revealed by considering how \(\Lcal_b^Y(x)\) alters the eigenvalues of the symmetry operators \(\Lcal^X_a\). It is useful to consider the geometry shown in \autoref{fig:punctured_plane}, where we have deformed the cylinder into a plane with a hole cut out. In this picture, it is clearer that the eigenvalues of \(\Lcal^X_a\) are revealing something about the topological charge (that is, the anyon content) of the edge. Indeed, if we transport a \(b\) anyon from far away and deposit it on the edge, we change the phase on \(\Lcal^X_a\) by the winding phase between \(a\) and \(b\):
            \begin{equation}
                \Lcal_b^Y(x) \Lcal^X_a = (R^{ab} R^{ba}) \Lcal^X_a \Lcal_b^Y(x),
                \label{eqn:op_winding}
            \end{equation}
            where we have dropped the dependence of \(R\) on the fusion channel to avoid further subscripts.

            In terms of anyons, condensing \(b\) at the edge alters the topological charge of the edge, \(a_{\mathrm{edge}} \mapsto b \times a_{\mathrm{edge}}\). In the symmetry language, \(\Lcal_b^Y(x)\) moves the edge between symmetry sectors, as revealed by \(\Lcal^X_a\). Combining these two interpretations, we see that symmetry sectors of the edge can be labeled by a topological charge---one can imagine the presence of an anyon in the hole of \autoref{fig:punctured_plane}, provided the anyon in question is condensed at the edge.

            The last case we consider is that of a gapless (possibly chiral) edge, described by a conformal field theory (CFT). This is similar to the case of \(b\) being condensed at the edge, in that \(\Lcal_b^Y(x)\) does generically have some action on the low energy degrees of freedom without exciting the bulk~\cite{Tong2016}. Indeed, it has been understood for a long time that bulk anyons can be related to primaries of the edge CFT through the action of \(\Lcal_b^Y(x)\), as we now briefly recall from \autoref{sec:background}. The restriction of \(\Lcal_b^Y(x)\) to the edge can be divided up into a linear combination of primary fields and their descendants, as is true of all operators in the CFT. The chiral algebra is the algebra of local operators, so \(\Lcal_b^Y(x)\) is only in the chiral algebra if \(b = 1\), as otherwise \(\Lcal_b^Y(x)\) is nonlocal---it involves a string which spans the entire length of the cylinder. We can label the primary field which contains \(\Lcal_b^Y(x)\) in its descendants by the anyon \(b\).

            It also remains true that \(\Lcal_b^Y(x)\) moves the edge between symmetry sectors---though, in the CFT context, the primary field content is usually of greater interest.

        \subsubsection{Bulk phase transitions and the second edge}
            \label{subsec:2nd_edge}

            Bulk anyon condensation can be studied within the framework developed above. However, one must be cautious in studying the edge in this context. Universal and nontrivial statements can be made about the edge only when it is protected by the bulk gap. As the bulk gap closes in the condensation phase transition, we may lose control of our description of the edge. We could recharacterise the edge after the transition is complete, but we would then necessarily be restricted to a purely algebraic description of the edge, based only on the anyon content of the bulk~\cite{Kitaev2012,Burnell2018} (without additional data from other sources~\cite{Kim2022,Kim2022modular}). Similarly, the usual bulk-boundary correspondence does not clearly apply during the transition, so CFT tools are also not clearly valid~\cite{Ginsparg1988,Francesco1997}

            To be able to keep track of the edge degrees of freedom all the way through the transition, we must be more specific about the condensation protocol. Suppose we wish to condense an anyon \(c\) in the bulk. (Recall that \(c\) must be a self boson.) Rather than tuning parameters in the bulk Hamiltonian uniformly, so that \(c\) is condensed everywhere, we instead tune parameters in the part of the system below the horizontal line \(y = y_c \gg \xi\). While the bulk gap closes in the lower part of the system \(y > y_c\), the gapped segment \(y_c > y > 0\) protects the edge degrees of freedom. A local perturbation at the edge would have to tunnel an excitation through the gapped segment, with its finite correlation length, in order to couple to the gapless part of the system. The amplitude for this process is exponentially small in \(y_c/\xi\).

            Once the condensation in the part of the system \(y > y_c\) has been accomplished, the domain wall between the condensed and uncondensed parts of the phase can be gapped (\autoref{sec:background})~\cite{Bais2009,Kitaev2011}. This is the eponymous second edge. Then, this domain wall can be moved towards the vacuum edge until \(\xi \gg y_c\). Once the second edge is within a correlation length of the first edge, it loses its distinct identity. The sequence of two edges should now be regarded as a single entity.

            The low energy degrees of freedom on the edge can still be sensibly identified as distinct to those belonging to the bulk because the bulk has no low energy excitations, even at \(y_c\) as the domain wall is gapped. This is the feature that required us to make the restriction to condensation phase transitions. If we considered an arbitrary phase transition, then the domain wall at \(y_c\) could be gapless. Then, this domain wall cannot be moved close to the first edge while retaining a sharp identification of degrees of freedom. The effect of the gapless second edge would have to be incorporated, which is beyond the scope of this work.

            The two edge model is pictured in the right hand side of \autoref{fig:cond_edge}. In the next section, we explore how this model reveals the consequences of anyon condensation for the edge.

    \subsection{General Consequences}
        \label{sec:consequences}

        The two edge model reveals that bulk anyon condensation is a \emph{symmetry breaking} transition of the edge (\autoref{subsec:symm_break})~\cite{Chatterjee2023}. Furthermore, when the edge is described by a CFT, it becomes straightforward to show that the chiral central charge (related to the thermal current on the edge) is invariant across the transition (\autoref{subsec:central_charge}). In contrast, purely algebraic descriptions are only capable of addressing the chiral central charge modulo \(8\)~\cite{Kitaev2006,Burnell2018}, though more detailed analyses of the bulk do show that that the chiral central charge is properly invariant~\cite[Eq.~(58)]{Kim2022modular}. Beyond the central charge, we can characterise how the primary field content of the edge is altered equally straightforwardly (\autoref{subsec:ext_chiral}). More sophisticated analyses also predict the primary field content, but require more mathematical machinery~\cite{Bais2009,Kong2018,Kong2020,Kong2021}.

        \subsubsection{Symmetry Breaking}
            \label{subsec:symm_break}

            The analogy between anyon condensation and symmetry breaking is already prevalent in the literature~\cite{Burnell2018,Chatterjee2023}. However, topological phases do not possess local order parameters, and so this analogy usually comes with several caveats. For the description of the edge, no such caveats are necessary. The transition is one of symmetry breaking in a more conventional sense than for the bulk. Indeed, we will see that the edge of the condensate even acquires a local order parameter~\cite{Lichtman2021}.

            The symmetry operators in question are the \(\Lcal^X_a\) strings of \autoref{sec:two_edge}. The order parameter is the vertical string operator \(\Lcal^Y_c(x)\) corresponding to the condensed anyon \(c\). We picture this in the two edge geometry. In the condensate phase, \(c\) is equivalent to the vacuum, so this string operator can start at the domain wall between the condensate and uncondensed phases (\autoref{fig:cond_edge}). Crucially, this means that \(\Lcal^Y_c(x)\) is a local operator when the two edges are made very close to each other. This means it is a candidate to serve as an order parameter for the edge model~\cite{Lichtman2021}.

            Further, \(\Lcal^Y_c(x)\) satisfies all the restrictions we demanded of the edge Hamiltonian. It does not excite the bulk, and is local. As such, it is a legitimate term that may appear in the effective edge Hamiltonian.

            Now consider some anyon \(a\) which becomes confined in the condensate phase, so that \(R^{ac}R^{ca} \neq 1\). The loop operator \(\Lcal^X_a\) no longer commutes with the condensate bulk Hamiltonian---it costs energy proportional to the circumference of the cylinder to implement~\cite{Trebst2007tension,Burnell2018}. Nor can it still be implemented in the narrow band of the original phase between the two edges, as \(\Lcal^Y_c(x)\) may now appear in the Hamiltonian. Combining Eq.~\eqref{eqn:op_winding} and \(R^{ac}R^{ca} \neq 1\), we see that \(\Lcal^X_a\) and \(\Lcal^Y_c(x)\) do not commute, so the presence of \(\Lcal^Y_c(x)\) in the edge Hamiltonian \emph{explicitly breaks} the symmetry \(\Lcal^X_a\).

            Similarly, a nonzero expectation value \(\avg{\Lcal^Y_c(x)}\) (generically expected when \(\Lcal^Y_c(x)\) can appear in the Hamiltonian) reveals the breaking of the symmetry \(\Lcal^X_a\). \(\Lcal^Y_c(x)\) indeed functions as an order parameter.

            We see that the confinement of an anyon \(a\) is reflected in the symmetry breaking of the edge model. The condensation of \(c\) also identifies any anyon species related by fusion of \(c\). This manifests as a quotient of the symmetry group by the subgroup generated by \(\Lcal^X_c\).

            The characterization of the edge in terms of symmetry and symmetry breaking allows us to relate condensation phase transitions to much more familiar physics. Further, the expression of the bulk phase transition on the edge mirrors the effect of condensation on the anyon content.

        \subsubsection{Invariance of the chiral central charge}
            \label{subsec:central_charge}

            A particular limitation of algebraic theories of anyons is that they are only sensitive to the \emph{topological central charge}, as opposed to the chiral central charge \(c_-\). It is the latter that characterises the CFT appearing at the edge of a chiral topological phase (when the edge can be described by a CFT), and relates to physically observable quantities. For instance, a small (compared to the bulk gap) but finite temperature \(T\) can create excitations in the low energy modes at the edge. A nonzero \(c_-\) indicates there is an excess of such modes which propagate in one direction, and leads to a net heat current at the edge (\(\hbar=k_B=1\))~\cite{Read2000,Kitaev2006},
            \begin{equation}
                I = \frac{\pi}{12} c_- T^2.
                \label{eqn:chiral_current}
            \end{equation}
            The topological central charge is equal to the chiral central charge modulo \(8\),
            \begin{equation}
                c_- \equiv c_-^{\mathrm{top}} \mod 8.
            \end{equation}
            It relates to braiding and fusion properties of the bulk anyons, but does not completely characterise the edge. Algebraic descriptions of anyons---that is, the description in terms of a unitary braided fusion category (UBFC)---miss this important physical feature.

            As such, while only studying the bulk anyons can, and does~\cite{Kitaev2006,Bais2009}, lead to the conclusion that the topological central charge does not change when the bulk goes through a condensation phase transition, such arguments do not show that the chiral central charge is also invariant. With the two edge model in hand, this result becomes transparent.

            Roughly, \(c_-\) counts the difference in the number of right- and left-moving modes in the CFT~\cite{Ginsparg1988,Francesco1997}. In order for \(c_-\) to change, other chiral modes must be available to scatter those of the edge CFT. In a generic phase transition, the closing of the bulk gap can provide such modes. In a condensation phase transition, the two edge construction shows that all the gapless modes can be kept separated from the edge.

            Considering both edges as a single entity, the fact that the domain wall to the condensate phase is gapped shows that it cannot contribute to \(c_-\). Thus \(c_-\) for the combined edge is due entirely to the edge between the intermediate uncondensed phase and the vacuum. That is, the chiral central charge for the edge of the condensate phase is that same as that for the uncondensed phase.

            \subsubsection{Alternative geometry}
                \label{subsubsec:alt_geom}

                \begin{figure}
                    \centering
                    \includegraphics[width=\linewidth]{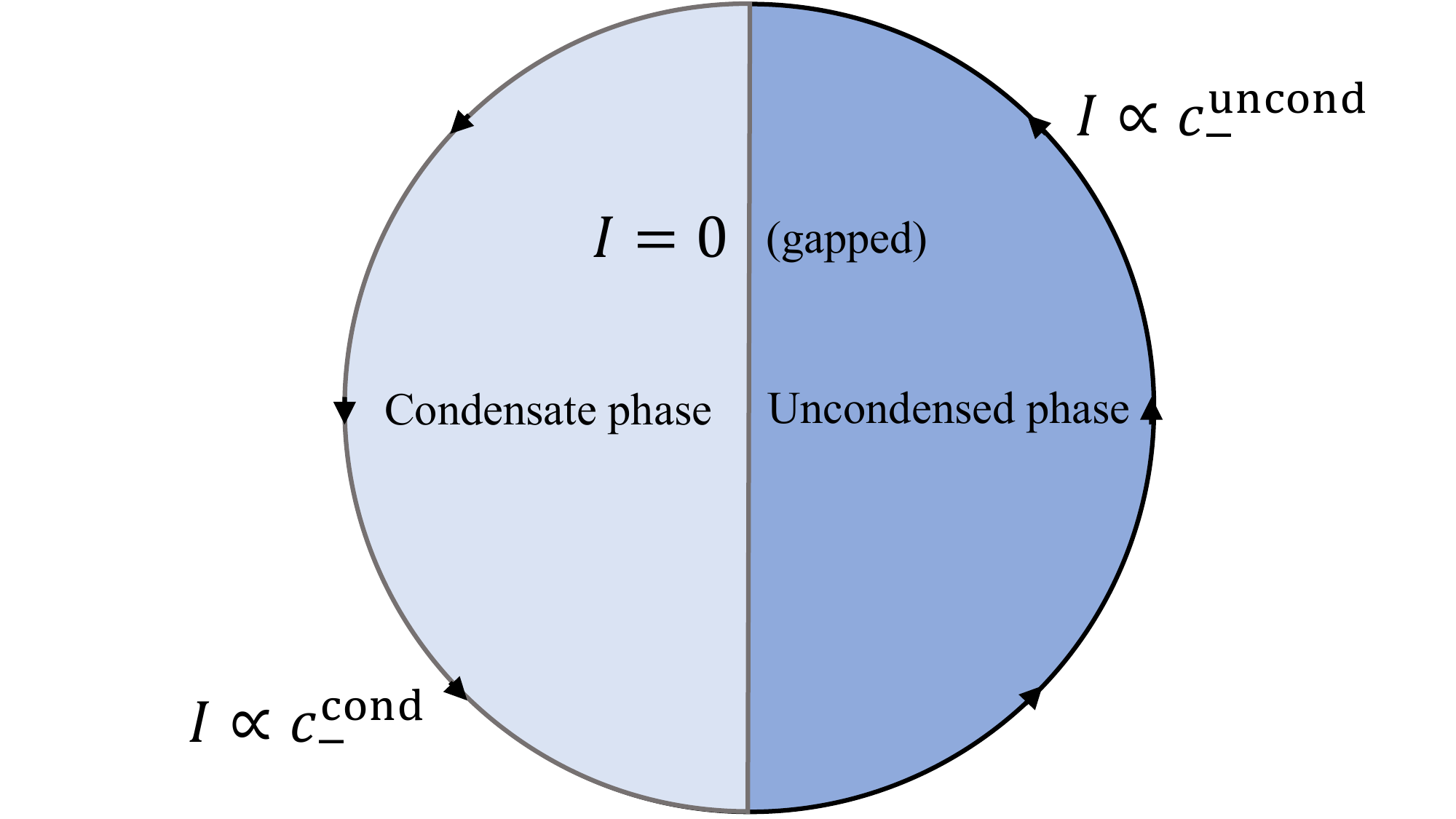}
                    \caption{\label{fig:disk}A disk geometry with one half of the system condensed shows that the chiral central charge of the edge is invariant through a condensation phase transition. At a temperature \(T\) much less than the bulk gap, the edge of each phase carries a current \(I = (\pi/12)c_- T^2\)~\cite{Read2000,Kitaev2006}. With the fact that domain walls between phases related by condensation can be gapped~\cite{Kitaev2012}, the conservation of energy then demands that the current along each half of the disk is equal, and thus that \(c_-\) is the same between each phase.}
                \end{figure}

                The two edge model uses a cylindrical geometry in order to access the symmetries \(\Lcal^X_a\). If we are not interested in symmetry properties, and only wish to conclude that the chiral central charge is invariant through as rapid a thought experiment as possible, a disk geometry is more useful.

                Consider the disk shown in \autoref{fig:disk}, where one half of the system is in a topological phase, and the other half is in the condensate phase. The low temperature energy current along each edge of the system is related to the chiral central charge by Eq.~\eqref{eqn:chiral_current}. The conservation of energy implies that the difference between the current along the edges of the condensate and uncondensed phases is given by the current along the domain wall.

                Recall from \autoref{sec:background} that domain walls between phases related by condensation phase transitions can be gapped. By adding additional perturbations to the wall, we may ensure that it is, indeed, gapped. (There is no loss of generality in assuming the domain wall is gapped, as the necessary perturbations cannot affect \(c_-\) in either bulk phase, as the perturbations are confined to the wall.) Thus the domain wall cannot carry any energy current at low temperatures. We conclude that the current along each external edge is equal, and thus that \(c_-\) between the two phases is also equal.

                This construction makes it particularly clear that the fact we are using to conclude invariance of the chiral central charge is the existence of gapped domain walls between phases related by condensation.

        \subsubsection{Extension of the chiral algebra}
            \label{subsec:ext_chiral}

            The chiral central charge is, in a sense, the coarsest information characterising a CFT. A clear picture of its behaviour is necessary, but not sufficient for a complete description of the edge. The two edge model also gives us a clean way to understand how anyon condensation affects the primary fields of the CFT, which furnishes this more complete description.

            The condensation of the anyon \(c\) extends the local algebra of the edge CFT---the \emph{chiral algebra}---by the primary labeled by \(c\) (\autoref{subsec:one_edge})~\cite{Francesco1997}. The two edge picture reveals what this means: the chiral algebra is the algebra of local operators, and in the two edge model \(\Lcal^Y_c(x)\) is local. As such, it now belongs to the chiral algebra.

            This simple observation leads to two major consequences for the other primary fields. The primaries that labeled chiral algebra equivalence classes (\autoref{sec:background}) should now be regarded modulo fusion by \(c\), as the chiral algebra now includes \(c\).

            Additionally, any primary field corresponding to a confined anyon \(b\) generically becomes gapped, and so disappears from the CFT describing the low energy degrees of freedom. As \(c\) is local, it can be added to the Hamiltonian. This gives an energetic cost to \(b \bar{b}\) creation operators: \(b\) acquires a mass.

            These two preceding points should be familiar. They are once again a reflection of the algebraic structure of condensation on the bulk model, identification and confinement, now appearing in the CFT primaries. The two edge model makes the process by which this occurs comprehensible.

\section{Achiral example: layers of the toric code}
    \label{sec:toric}

    Section~\ref{sec:general} outlined the general construction we have developed to study the effect of condensation transitions on the edge of a topological phase. In this section, we make this abstract procedure concrete through a close examination of a simple example.

    The toric code is the most widely studied model of topological order in the literature at this time~\cite{Kitaev2003,Dennis2002,Kitaev2009,Wen2017}. It has many attractive properties: it functions as an error correcting code for the storage of quantum information~\cite{Kitaev2003,Dennis2002}; it is a paradigmatic example of a spin liquid~\cite{Wen2017}; and it is exactly soluble, making very detailed analyses possible. Relevant for our purposes, it also admits a condensation transition~\cite{Vidal2009,Vidal2009bound,Dusuel2011perturbed}, as does the multi-layer toric code~\cite{Wiedmann2020bilayerTC,Schamriss2022layerTC}.

    Implementing the condensation transition moves the model away from exact solubility~\cite{Tupitsyn2010}, but the two edge model of \autoref{sec:general} remains soluble in a particular limit. This allows us to inspect the effects of anyon condensation on the edge in very explicit terms.

    In \autoref{sec:TC_edges} we review a construction of the toric code that is particularly convenient for the study of its edges~\cite{Ho2015}. Indeed, the effective model for the edge in this construction is just the transverse field Ising model, which is also exactly soluble and, as predicted, possesses a global (spin flip) symmetry. We then proceed to an analysis of the effects of condensation on the edge model (\autoref{sec:TC_cond}). We will find that condensing an anyon in the bulk introduces terms on the edge that appear as a longitudinal field in the effective Hamiltonian, explicitly breaking the spin flip symmetry of the original model. This is as predicted in \autoref{sec:general}. Lastly, we extend our analysis to an arbitrary anyon condensation in a stack of multiple layers of the toric code, and show that this more general scenario actually reduces to the single-layer case (\autoref{sec:TC_multi}).

    \subsection{The toric code and its edges}
        \label{sec:TC_edges}

        The toric code~\cite{Kitaev2003,Dennis2002,Kitaev2009,Wen2017} hosts Abelian anyons, two of which are bosonic, and so admit condensation~\cite{Burnell2018}. Many expositions on, and studies of, the toric code have been made in the literature. A study of the edge of the toric code is most easily facilitated by a less standard construction developed in Ref.~\cite{Ho2015}. This utilizes the \emph{Wen plaquette} model for the toric code phase~\cite{Wen2003}, which is unitarily related to the usual toric code model~\cite{Kitaev2003}. In this section, we briefly review this model.

        \subsubsection{Bulk}
            \label{subsec:TC_bulk}

\begin{figure}
                \centering
                \includegraphics[width=\linewidth]{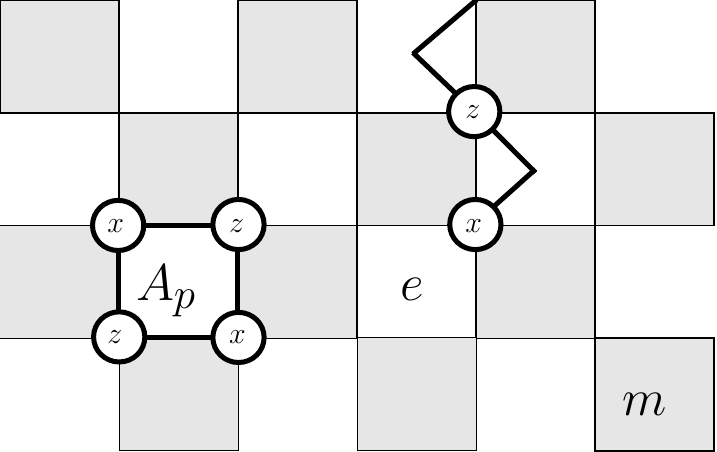}
                \caption{\label{fig:wen_lattice}\emph{The Wen plaquette model.} Qubits are placed on the vertices of a square lattice, and an operator \(A_p\) is associated to each plaquette \(p\). The Hamiltonian of the system is \(H = - K\sum A_p\) (with \(K>0\)). The system has anyons appearing as \(-1\) eigenvalues for some \(A_p\) operator, and these anyons can be moved for no energy cost with the appropriate string-like operators. The anyon species on the white plaquettes as shown is denoted \(e\), and that on the shaded plaquettes is \(m\).}
            \end{figure}

            The bulk of the Wen plaquette model~\cite{Wen2003} is pictured in \autoref{fig:wen_lattice}. It places qubit degrees of freedom on the vertices of a square lattice, and defines \emph{plaquette operators} associated to each face \(p\):
            \begin{equation}
                A_p = \begin{array}{c c}\sigma^x_2 & \sigma^z_1 \\ \sigma^z_3 & \sigma^x_4 \end{array}.
            \end{equation}
            Here, qubits are labelled anticlockwise around the plaquette, and have been written so as to reflect their geometry.

            The Hamiltonian is the negative sum of all the plaquette operators,
            \begin{equation}
                H = - K \sum_{p} A_p,
            \end{equation}
            where we take \(K > 0\). Crucially, all the plaquette operators \(A_p\) commute~\cite{Wen2003}.

            Thus the eigenstates of \(H\) can be identified as the simultaneous eigenstates of each plaquette operator. The ground space of the model is the shared \((+1)\)-eigenspace of all the plaquette operators. Excitations above this state are revealed as \((-1)\)-eigenvalues of some \(A_p\). This lets us identify such an excitation as belonging to a plaquette of the lattice.

            These excitations can be generated in pairs, and subsequently moved around for zero energy cost. Indeed, acting by \(\sigma^x\) (\(\sigma^z\)) at a vertex anticommutes with the \(A_p\) operators to the north-east and south-west (north-west and south-east) of that vertex, flipping their sign. This produces two excitations, both on plaquettes with the same shading in the checkerboard pattern of \autoref{fig:wen_lattice}. By flipping more pairs of plaquettes, the two excitations can be moved around, still on the same colour plaquettes. We identify these mobile excitations with distinct anyon species, \(e\) and \(m\), depending on the colour of the plaquette they live on. Say, \(e\) being based on the white plaquettes, and \(m\) on the shaded plaquettes. The combination of an adjacent \(e\) and \(m\) will be called \(\epsilon\).

            These excitations function as anyons. Creating a pair of \(e\) particles and wrapping one around a small closed loop produces a string operator which is the product of the \(A_p\) belonging to shaded plaquettes within that loop. Thus if an \(m\) anyon is present in that region, this operation gives a \((-1)\) phase to the wave function. In the notation of \autoref{sec:background}, \(R^{em}_\epsilon R^{me}_\epsilon = -1\). The \(A_p\) operators can be interpreted as loops of one kind of anyon or the other, which detect the presence of the other species.

        \subsubsection{Edge}
            \label{subsec:TC_boundary}

            \begin{figure}
                \centering
                \includegraphics[width=\linewidth]{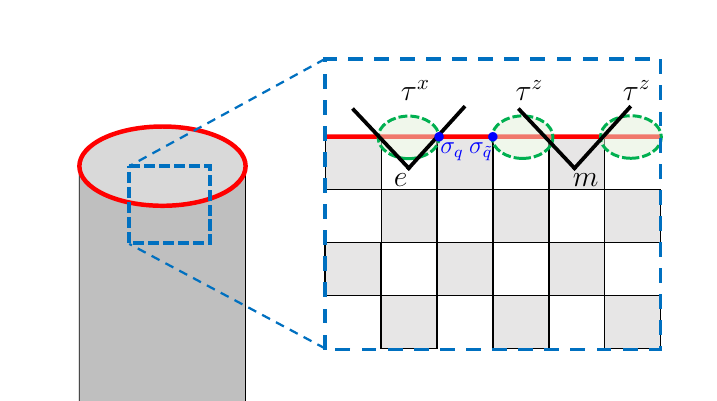}
                \caption{\label{fig:wen_edge}The Wen plaquette model on a semi-infinite cylinder. We place the edge of the system at the extreme north of the cylinder (up the page as shown). Each pair of white and shaded plaquettes on the edge has a qubit degree of freedom associated to it in the analysis of Ref.~\cite{Ho2015} (dashed green circles), and the small anyon movement operators act as Pauli operators on these degrees of freedom. An \(e\) operator \(S_{\tilde{q}}\) acts as the Pauli \(x\) on one such degree of freedom, which we write as \(\tau^x_{q}\) so as not to be confused with the Pauli operators on the physical qubits, \(\sigma_{q(\tilde{q})}^\alpha\). An \(m\) operator \(S_{q}\) acts as \(\tau^z_{q-1} \tau^z_q\) on adjacent degrees of freedom.}
            \end{figure}

            In Ref.~\cite{Ho2015}, a characterisation of the edge degrees of freedom of the Wen plaquette model in terms of its anyon content was deduced. The model considered is a single layer of the Wen plaquette model on a semi-infinite cylinder, such that there is an even number, \(L\), of plaquettes around the circumference of the cylinder. This allows us to unambiguously colour the plaquettes as white or shaded. We choose the terminating edge of the cylinder to be to the extreme north (\autoref{fig:wen_edge}).

            We label a shaded plaquette adjacent to the edge of this system \(q\), and write the Pauli operators which act on the edge vertices north-west (north-east) of this plaquette \(\sigma_q^\alpha\) (\(\sigma_{\tilde{q}}^\alpha\)) (\autoref{fig:wen_edge}).

            In Ref.~\cite{Ho2015}, it is shown that
            \begin{equation}
                \left\{ A_p, S_{\tilde{q}} = \sigma^z_{\tilde{q}} \sigma^x_{q+1}, \Lcal^X_m \right\}_{p,q},
                \label{eqn:bulk_dof_one}
            \end{equation}
            is a complete set of commuting operators for this system. Here, \(\Lcal^X_m\) is a path operator for the \(m\) anyons which wraps around the cylinder once. Each of these terms has an interpretation as an anyon movement operator---each \(A_p\) can be thought of as small loop operators for either \(e\) or \(m\) anyons, and we defined \(\Lcal^X_m\) as a loop operator explicitly, while the \(S_{\tilde{q}}\) operators can be interpreted as moving an \(e\) anyon from off the edge onto site \(\tilde{q}\), and then back off the edge through the other corner (\autoref{fig:wen_edge}).

            Additionally, the degrees of freedom belonging to the edge may be characterised. If the circumference of the cylinder is \(L \in 2\Z\) and the system is in a \((+1)\)-eigenstate of all of \(A_p\) and \(\Lcal^X_m\), then there are \(L/2\) qubit degrees of freedom remaining, which we associate with the edge. We can arrange these so that one is associated to every pair of shaded and white plaquettes \(q\) and \(\tilde{q}\) on the edge. We say that the edge qubit associated to \((q,\tilde{q})\) is located at edge site \(q\). We label the Pauli operators acting on these edge qubits by \(\tau_q^\alpha\) to distinguish them from the bulk Pauli operators.

            We could choose to fix \(\Lcal^X_m\) in its \((-1)\)-eigenstate, as it does not actually appear in the Hamiltonian. This will later lead to an antiperiodic edge theory, rather than a periodic one. In the language of \autoref{sec:general}, this is working in a different symmetry sector of \(\Lcal^X_m\).

            The relationship between the bulk degrees of freedom and the edge degrees of freedom is provided by the anyons. The path operators that take \(e\) and \(m\) anyons off the edge and then back on, \(S_{\tilde{q}} = \sigma^z_{\tilde{q}} \sigma^x_{q+1}\) and \(S_q = \sigma^z_{q} \sigma^x_{\tilde{q}}\) respectively, commute with the bulk Hamiltonian and \(\Lcal^X_m\), and can thus be interpreted as acting on the edge degrees of freedom. Any representation of \(S_{\tilde{q}}\) and \(S_q\) on the \(L/2\) edge qubits must preserve the (anti)commutation relationships between them. A possible choice is (\autoref{fig:wen_edge})~\cite{Ho2015}
            \begin{equation}
                S_{\tilde{q}} \mapsto \tau_q^x, \quad \text{and}\quad S_q \mapsto \tau^z_{q-1} \tau^z_q.
            \end{equation}

            A natural edge Hamiltonian for the Wen plaquette model, generated by some perturbations to the exact model, then has the form
            \begin{equation}
                H_\mathrm{e} =  - h\sum_q \tau^x_q -J \sum_q \tau^z_{q-1} \tau^z_q ,
                \label{eqn:H_edge}
            \end{equation}
            where \(J>0\) and \(h\) are constants. This is the transverse field Ising model (TFIM) in one dimension and with periodic boundary conditions. If we fixed \(\avg{\Lcal^X_m} = -1\), then one of the \(\tau^z_{q-1} \tau^z_q\) terms should have its sign reversed to enforce antiperiodic boundary conditions.

            The TFIM has a global \(\Z_2\) symmetry, corresponding to the operator \(\Lcal^X_e = \prod_q \tau^x_q\), which applies the Pauli \(x\) operator to every edge qubit and commutes with \(H_\mathrm{e}\). If we only allow perturbations to the bulk that maintain the bulk ground space, then all the resulting edge perturbations respect this symmetry. In this symmetric regime, this model has two distinct phases, one corresponding to large \(h\) (paramagnetic), and the other to large \(J\) (ferromagnetic). The phases are separated by a gapless critical point.

            Of course, this reflects the general understanding we developed in \autoref{sec:toric}. Requiring that the edge model not excite the bulk gap \emph{demands} the presence of a symmetry for each anyon of the bulk model. We explicitly constructed the edge Hamiltonian in blocks labeled by \(\avg{\Lcal^X_m}\), and discovered a model in each block with an \(\Lcal^X_e\) symmetry.

    \subsection{Condensation and edges of the toric code}
        \label{sec:TC_cond}

        The anyons \(e\) and \(m\) of the toric code are both bosons~\cite{Kitaev2003}, and so they can be condensed. The toric code is not exactly soluble all the way through the condensation transition~\cite{Tupitsyn2010}, but both extremes of the phase diagram are soluble. This makes a two edge model as described in \autoref{sec:general} particularly useful, as it retains some amount of solubility.

        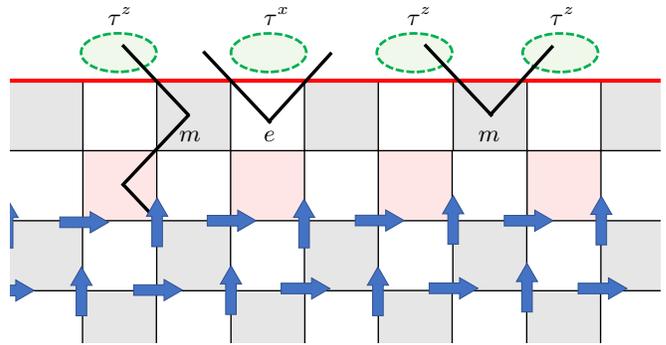
\begin{figure}
            \centering
            \begin{overpic}[width=\linewidth]{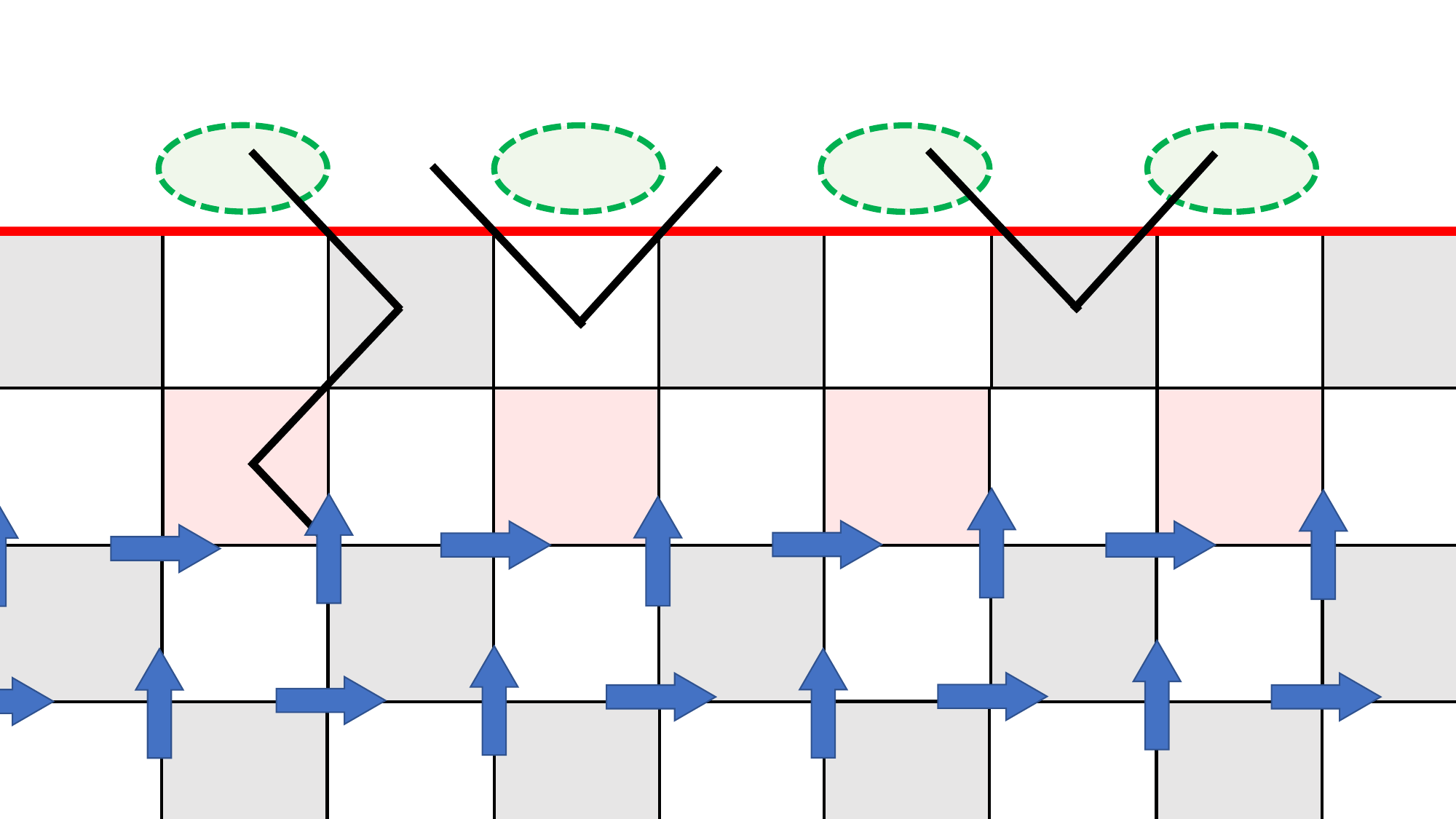}
                \put(26,31){\(m\)}
                \put(39,31){\(e\)}
                \put(72,31){\(m\)}

                \put(15,49){\(\tau^z\)}
                \put(39,49){\(\tau^x\)}
                \put(61,49){\(\tau^z\)}
                \put(83,49){\(\tau^z\)}
            \end{overpic}
            \caption{\label{fig:TC_2edge}The two edge model for the Wen plaquette. The Hamiltonian in the lower portion of the system, in the condensate phase, consists of local fields on the qubits (blue arrows). In the strip between the condensate phase and the edge, the Hamiltonian is that of the usual Wen plaquette, except that the plaquette operators that do not commute with the local fields are excluded (pink plaquettes). This leads to a \emph{smooth edge}. \(m\) anyons can be created for free from the domain wall and brought off the edge. These short strings act as \(\tau^z\) on the effective edge degrees of freedom.}
        \end{figure}

        We will focus on condensing \(m\). Microscopically, this is accomplished by adding pair-creation operators for \(m\) to the Hamiltonian, and making them the dominant term~\cite{Tupitsyn2010}. We parameterise the condensation Hamiltonian as
        \begin{equation}
            H_{\mathrm{c}} = -K\left[ \lambda \sum_p A_p + (1-\lambda) \sum_{p\;\mathrm{shaded}} \sigma^z_{p,\mathrm{NW}} + \sigma^x_{p,\mathrm{NE}} \right].
            \label{eqn:TC_Hcond}
        \end{equation}
        Here, \(\sigma^\alpha_{p,\mathrm{NW}}\) is the Pauli \(\alpha\) operator on the qubit in the north-west corner of the plaquette \(p\), and similarly for \(\sigma^\alpha_{p,\mathrm{NE}}\) in the north-east. The dimensionless parameter \(\lambda\) tunes the Hamiltonian between the Wen plaquette Hamiltonian \(\lambda = 1\) and a sum of local fields \(\lambda = 0\), for which the ground state is a product state. The latter is the extreme limit of the \(m\) condensate phase. The condensation transition is a gap closing transition at some \(\lambda_c\) between 0 and 1.

        At all \(\lambda \in (0,1)\), the Hamiltonian \eqref{eqn:TC_Hcond} has no known solution. The general analysis of \autoref{sec:general} does not rely on any kind of solubility, but in the interests of having a concrete calculation verifying those claims, we use the two edge geometry of \autoref{sec:general}.

        We choose some horizontal line \(y=y_c\) below which we will set \(\lambda=1\), and above which we will set \(\lambda = 0\). In the bulk of either region, the Hamiltonian is a sum of commuting terms. The ground state is a product state for \(y > y_c\) and the toric code ground state in the strip \(y_c > y > 0\) (measuring \(y\) increasing southwards, as in \autoref{fig:cond_edge}). At the interface of the two phases, the two Hamiltonians do not commute. If we demand that the condensate phase have the larger gap, then an effective model of the low energy degrees of freedom favours the formation of the product state below \(y=y_c\). The effective Hamiltonian is obtained by replacing Pauli operators on the qubits in a product state by their expectation values. The result is just a deletion of those \(A_p\) which anticommute with the local fields at the domain wall---those belonging to shaded plaquettes, which measure the presence of \(m\) anyons (\autoref{fig:TC_2edge}).

        Experts will recognise this structure as a \emph{smooth edge} of the toric code. In this case, where the condensate phase is a trivial product state, the two edge model reduces to a very thin strip of the toric code.

        To find an effective edge model, we use the same parameterisation of the edge degrees of freedom as in \autoref{sec:TC_edges}, so that the short anyon movement operators across the edge produce a TFIM in the effective edge model. First, we observe that the loop operator \(\Lcal^X_m\) must act trivially, as it can be absorbed it into the condensate part of the system, where it fixes the ground state. There is no longer a \(\avg{\Lcal^X_m} = -1\) sector.

        Next, we have additional short \(m\) movement operators which leave no excitations in the bulk---those which generate an \(m\) anyon from the domain wall and send it off the northern edge. These are the \(\Lcal^Y_m(x)\) operators we encountered in \autoref{sec:general}. Inspecting commutation relations between these strings and \(S_q\) and \(S_{\tilde{q}}\) from \autoref{subsec:TC_boundary}, we see that 
        \begin{equation}
            \Lcal^Y_m(x) \mapsto \tau^z_q
        \end{equation}
        acts as a longitudinal field.

        As a result, the model for the edge becomes
        \begin{equation}
            H_{\mathrm{c,e}} = - h\sum_q \tau^x_q -J \sum_q \tau^z_{q-1} \tau^z_q - k\sum_q \tau^z_q.
            \label{eqn:H_edgecond}
        \end{equation}
        This is the transverse field Ising model with an additional longitudinal field---the mixed field Ising model (MFIM).

        \subsubsection{Symmetry breaking}

            The most striking change to the edge model from that before condensation~\eqref{eqn:H_edge} is the breaking of the \(\Z_2\) symmetry. In the \(\tau\) degrees of freedom, the \(\Lcal^X_e\) symmetry acts as \(\prod_q \tau^x_q\). When \(k\neq0\) in Eq.~\eqref{eqn:H_edgecond}, this symmetry is explicitly broken.

            This has very significant changes on the phenomenology of the edge. The transverse field Ising model, with its symmetry imposed, has two distinct phases, separated by a gap closing transition at \(h/J=1\). The longitudinal field Ising model does not. It has a single gapped phase with a gapless \emph{point} at \(h/J=1\), \(k=0\). It also has a line \(h/J < 1\), \(k=0\), across which there is a first-order phase transition, but it is possible to interpolate between the ferromagnetic and paramagnetic phases separated by this line by moving through the \(h/J>1\), \(k \neq 0\) region.

            As such, the entire phase diagram for the edge is now in the same phase as the vacuum---it is trivial. This is expected. After condensation, the bulk enters the trivial product state phase.

        \subsubsection{Invariance of the chiral central charge}

            The predictions we had for the CFT describing the gapless degrees of freedom at the edge are largely trivial in this example. We describe the conclusions that can be made about the gapless point in the phase diagram of the edge for completeness.

            The CFT describing the gapless point \(h/J=1\) in the TFIM is the Ising CFT. It is closely related to the free Majorana fermion CFT~\cite{Ginsparg1988,Francesco1997}.

            The Ising CFT has a chiral central charge of \(c_- = 0\). This is not to be confused with the holomorphic or antiholomorphic central charges, \(c\) and \(\bar{c}\), that describe the number of right- and left-moving degrees of freedom. The Ising CFT has \(c = \bar{c} = 1/2\).

            Once the longitudinal field gets turned on after condensation, all the degrees of freedom become gapped, and the CFT describing the low energy degrees of freedom is the zero CFT. This also has \(c_- = 0\).

            This is typical of achiral phases. There is in general no anomaly which protects the gapless modes on the edge in this case (the anomaly in question usually being \(c_-\)). As such, the edges of such phases are frequently gapped, regardless of any condensation in the bulk.

        \subsubsection{Extension of the chiral algebra}

            The effect of condensation on the primary fields at the gapless point is also straightforward. The Ising CFT has three relevant primary fields with equal holomorphic and anti-holomorphic conformal weights, usually denoted \(\epsilon\) (the energy), \(s\) (spin) and \(\mu\) (disorder). (These can each be decomposed into purely left- or right-moving parts, none of which are in the chiral algebra.)

            Our colloquial definition of the chiral algebra is that it is the algebra of local operators. Prior to condensation, only the energy \(\epsilon\) is simultaneously local, relevant, and respects the symmetry. Microscopically, it represents the coupling \(h-J\) in the Hamiltonian, and measures deviation from the critical point.

            The spin \(s\) is local and relevant, but does not respect the \(\Z_2\) symmetry. It is dual to the disorder \(\mu\), which is nonlocal and does not respect the dual symmetry which was due to \(\Lcal^X_m\) in our construction. After condensation, we add \(s\) to the list of permissible perturbations from criticality. The fields \(s\) and \(\mu\) have nontrivial braiding, so usually this would imply that \(\mu\) gets deleted from the CFT. However, the CFT becomes trivial once deformed by \(s\), so this is an uninteresting statement.

    \subsection{Condensation in multiple layers of the toric code}
        \label{sec:TC_multi}

        Many interesting topological phases can be constructed from multiple layers of the toric code~\cite{Bombin2006,Wang2013}. Often, such constructions involve condensing composite anyons which span several layers~\cite{Wang2013}. As such, it is useful to generalise our discussion of condensation in the toric code, and its effect on the edge, to the multi-layer case.

        We will first summarise a very simple case, where condensation occurs separately in each layer (\autoref{subsec:separate}). Then, we show how to reduce the general case to the simple one, and enumerate a few examples (\autoref{subsec:TC_multi_general}).

        \subsubsection{Separate condensation}
            \label{subsec:separate}

            Consider a stack of \(N\) layers of the Wen plaquette model for the toric code, wrapped into the semi-infinite cylinder geometry. A complete set of commuting observables for the entire stack can be obtained by concatenating such sets for each layer,
            \begin{equation}
                \left\{ A_p^l, S_{\tilde{q}}^l = \sigma^{z,l}_{\tilde{q}} \sigma^{x,l}_{q+1}, \Lcal_m^{X,l} \right\}_{p,q,l},
                \label{eqn:bulk_dof_many}
            \end{equation}
            where a superscript \(l\) indexes the layers. Similarly, the effective model for the edge is just \(N\) copies of the effective model for one edge, the TFIM,
            \begin{equation}
                H_\mathrm{e} =  -\sum_{l=0}^{N-1}\left(h\sum_q \tau^x_q +J \sum_q \tau^z_{q-1} \tau^z_q \right).
                \label{eqn:H_edge_multi}
            \end{equation}

            Within the two edge construction, condensing an \(m^l\) anyon on layer \(l\) produces a new effective model for the edge which has an additional longitudinal field on layer \(l\) (\autoref{sec:TC_cond}). Condensing multiple \(m^{l_i}\) anyons, for \(i \in \{0,\ldots,n-1\}\), introduces a longitudinal field on each of the \(n\) layers \(l_i\),
            \begin{equation}
                H_\mathrm{c,e} =  -\sum_{l=0}^{N-1}\left(h\sum_q \tau^{x,l}_q +J \sum_q \tau^{z,l}_{q-1} \tau^{z,l}_q \right) - k \sum_{i=0}^{n-1}\sum_q \tau^{z,l_i}_q .
                \label{eqn:H_edge_cond_multi}
            \end{equation}
            The layers \(l_i\) are now described by MFIMs.

            The effect of condensation on the symmetry structure can similarly be treated separately in each layer. Before condensation, the separate layers each have two independent symmetry generators, \(\Lcal^{X,l}_e\) and \(\Lcal^{X,l}_m\), making the total symmetry group \(\Z_2^{2 N}\). After condensation, \(\Lcal^{X,l}_e\) become confined and \(\Lcal^{X,l}_m\) acts trivially---they both disappear from the symmetry representation on the edge degrees of freedom. The new symmetry group is \(\Z_2^{2 (N-n)}\).

        \subsubsection{General case}
            \label{subsec:TC_multi_general}

            Our choice of operators that define the degrees of freedom on the edge was well suited to the particular condensation above, but is not unique. To address general anyon condensations, we tailor the choice of degrees of freedom to the particular condensation of interest. In particular, if \(n\) anyons \(\{a^{0}, \ldots, a^{n-1}\}\) are condensed, we choose degrees of freedom \(\tau^{\alpha,l}_q\) such that the \(\Lcal^Y_{a^i}(x)\) term in the effective edge model is still a longitudinal field. This means that the rest of the edge model will no longer be a TFIM~\footnote{We could have made the alternative choice to preserve the TFIM part of the edge model, in which case the longitudinal field would change form. The result of either choice is related to the other by a unitary transformation.}.

            Such a choice is possible due to the fact that in order to condense \(\{a^{0}, \ldots, a^{n-1}\}\), they must all be self and mutual bosons, just as \(\{m^{l_0},\ldots,m^{l_n}\}\) were. It is possible to find anyons \(\{a^{n},\ldots,a^{N-1}\}\) and \(\{b^0,\ldots,b^{N-1}\}\) such that each \(a^i\) anyon is a self boson, a mutual semion with \(b^i\), and a mutual boson with every other anyon in the list. Furthermore, the full set of \(a\) and \(b\) anyons should be independent under fusion.

            It is useful to describe what the bosonic condition means when viewing each \(a^i\) as a composite anyon
            \begin{equation}
                a^i = \prod_{l=0}^{N-1} c^l_i,
                \quad\text{where}\quad
                c^l \in\{1,e^l,m^l, \epsilon^l = e^l m^l\}.
            \end{equation}
            We can view \(a^i\) as a vector in \((\Z_2^2)^N\) by identifying \(e^l\) and \(m^l\) as basis vectors in this space, and taking fusion of anyons to be addition of vectors. It is convenient to split \(a^i\) into a part composed of \(e^l\) anyons, \(a^i_e \in \Z_2^N\), and a part composed of \(m^l\) anyons, \(a^i_m \in \Z_2^N\), and writing \(a^i = a^i_e \oplus a^i_m\). Then, the statement that \(a^i\) is a self-boson is just
            \begin{equation}
                \avg{a^i_e, a^i_m} = 0,
            \end{equation}
            where \(\avg{\cdot,\cdot}\) is the usual \(\Z_2\) valued dot product in \(\Z_2^N\). The \(e\) and \(m\) parts of \(a^i\) must be orthogonal in \(\Z_2^N\).

            The statement that \(a^i\) and \(a^j\) are mutual bosons may be expressed as the vanishing of a so called \emph{symplectic} bilinear form, which is expressed in terms of the usual dot product as
            \begin{equation}
                (a^i, a^j) := \avg{a^i_e, a^j_m} + \avg{a^i_m, a^j_e} = 0.
                \label{eqn:symplectic}
            \end{equation}

            Thus the condensed anyons \(\{a^{0}, \ldots, a^{n-1}\}\) furnish a linearly independent set of vectors in \((\Z_2^2)^N\), that are orthogonal with respect to the bilinear form~\eqref{eqn:symplectic}. It is possible to complete this basis and obtain a set
            \begin{equation}
                \{a^{0}, \ldots, a^{N-1}, b^0, \ldots b^{N-1}\}
                \label{eqn:basis_anyons}
            \end{equation}
            such that \(\avg{a^i_e, a^i_m} = \avg{b^i_e, b^i_m} = 0\) and \((a^i,a^j) = (b^i,b^j) = 0\) while \((a^i,b^j) = \delta^{ij}\). The problem of finding such a basis also occurs in classical mechanics, when finding a complete set of conjugate coordinates and momenta; and in quantum error correction, when finding logical operators and primary errors for a given set of stabilisers~\cite{Gottesman2009}.

            Translating back into the anyon language, we have found a set of anyons that have \emph{the same} fusion and braiding relations as \(\{m^{0}, \ldots, m^{N-1}, e^0, \ldots e^{N-1}\}\). Each anyon fuses with itself to give the vacuum, and braiding \(a^i\) with \(b^i\) gives a \((-1)\) sign. All other relations are trivial, or determined by the anyon theory being Abelian. Then the effect of condensing \(\{a^{0}, \ldots, a^{n-1}\}\) is easy to characterise. Each of \(\{b^{0}, \ldots, b^{n-1}\}\) becomes confined, while \(\{a^{n}, \ldots, a^{N-1}, b^n, \ldots b^{N-1}\}\) are unaffected. The algebra here is the same as in \autoref{subsec:separate}.

            We define edge operators \(\tau_q^{z,l_i}\) as the short strings \(\Lcal^Y_{a^i}(x_q)\), where \(x_q\) is the position of the edge plaquette \(q\). The conjugate \(\tau_q^{x,l_i}\) are associated to short string operators of \(b^i\) which cross one \(\Lcal^Y_{a^i}(x_q)\), so that \(\tau_q^{z,l_i}\) and \(\tau_q^{x,l_i}\) anticommute, but commute with \(\tau\) operators at other sites. The natural physical Hamiltonian in the thin uncondensed cylinder of the two edge model still consists of the \(e^l\) and \(m^l\) anyon movement operators \(S^l_{\tilde{q}}\) and \(S^l_q\). The effective \(\tau\) operators corresponding to the physical string operators can be deduced through the commutation relations between the \(S^l_{\tilde{q} (q)}\) strings and the \(a^i\) and \(b^i\) strings.

            With this convention, the terms added by condensation will always be a longitudinal field,
            \begin{equation}
                -V_{n} := -\sum_{i=0}^{n-1} \sum_q \tau^{z,l_i}_q,
            \end{equation}
            where we set \(k = 1\).

            In the original lattice model, the edge model is always a sum of \(S_q\) and \(S_{\tilde{q}}\) operators. The expression of these operators as products of \(\tau^\alpha_q\) operators will depend on the choice of basis \(\{a^{0}, \ldots, a^{N-1}, b^0, \ldots b^{N-1}\}\). Nonetheless, symmetry breaking will always appear as reducing \(\Z_2^{2 N}\) down to \(\Z_2^{2 (N-n)}\).

            We enumerate a few examples below to make this construction clear.

        \subsubsection{Condense \(m^0 m^1\) in \(N=2\) layers}

            We choose \(a^0 = m^0 m^1\), \(a^1 = m^1\), \(b^0 = e^0\), \(b^1 = e^0 e^1\). Taking the physical Hamiltonian to be a sum of \(S^l_{\tilde{q}}\) and \(S^l_q\) operators, and inspecting the braiding relations between the \(e^l\) and \(m^l\) anyons defining those strings with the \(a^l\) and \(b^l\) strings defining the \(\tau\) operators, we find an edge Hamiltonian,
            \begin{align}
                H_\mathrm{c,e} = - V_1 - & \sum\left( J_0 \begin{smallmatrix} \tau^z & \tau^z \\ \tau^z & \tau^z \end{smallmatrix} + h_1 \begin{smallmatrix}\tau^x \\ 1 \end{smallmatrix}\right) \nonumber \\
                - &\sum\left( J_1 \begin{smallmatrix} 1 & 1 \\ \tau^z & \tau^z \end{smallmatrix} + h_2 \begin{smallmatrix}\tau^x \\ \tau^x \end{smallmatrix}\right).
                \label{eqn:mm}
            \end{align}
            We have omitted all qubit specifying subscripts, and indicated the positions of operators graphically.

            In the limit \(J_l, h_l \ll 1\), corresponding to a large amplitude for tunneling \(m^0 m^1\) from the condensate to the edge, perturbation theory on the degenerate \(V_1\) ground space gives the Hamiltonian
            \begin{equation}
                H_\mathrm{c,e} \approx -V_1 - (J_0 + J_1)\sum \begin{smallmatrix} 1 & 1 \\ \tau^z & \tau^z \end{smallmatrix} - h_1 h_2 \sum \begin{smallmatrix}1 \\ \tau^x \end{smallmatrix} + \cdots.
            \end{equation}
            Here, we have assumed all parameters are positive, and we have replaced \(\begin{smallmatrix} \tau^z & \tau^z \\ \tau^z & \tau^z \end{smallmatrix}\) with \(\begin{smallmatrix} 1 & 1 \\ \tau^z & \tau^z \end{smallmatrix}\) as these have the same action on the \(V_1\) ground space.

            The result is a trivial theory on the top layer and a TFIM on the second layer with transformed parameters. We expect a phase transition when \(J_0 + J_1 = h_0 h_1\). In the special case of identical layers (\(J_0 = J_1 = J\) and \(h_0 = h_1 = h\)) this is \(2 J = h^2\).

            When \(J_l, h_l \gg 1\), we may safely neglect the \(V_1\) term, and this becomes two copies of the transverse field Ising model, with a phase transition at \(h_0/J_0=1\) and \(h_1/J_1 = 1\).

        \subsubsection{Condense \(m^0 m^1 m^2\) in \(N=3\) layers}

            We choose \(a^0 = m^0 m^1 m^2\), \(a^1 = m^0\), \(a^2 = m^2\), \(b^0 = e^1\), \(b^1 = e^0 e^1\), and \(b^2 = e^1 e^2\), resulting in an edge Hamiltonian
            \begin{align}
                H_\mathrm{c,e} = - V_1 - & \sum\left(J_0 \begin{smallmatrix} 1 & 1 \\ \tau^z & \tau^z \\ 1 & 1 \end{smallmatrix} + h_0 \begin{smallmatrix}\tau^x \\ \tau^x \\ 1 \end{smallmatrix}\right) \nonumber \\
                - &\sum\left(J_1 \begin{smallmatrix} \tau^z & \tau^z \\ \tau^z & \tau^z \\ \tau^z & \tau^z \end{smallmatrix} + h_1 \begin{smallmatrix}\tau^x \\ 1 \\ 1 \end{smallmatrix}\right) \nonumber \\
                - &\sum\left(J_2 \begin{smallmatrix} 1 & 1 \\ 1 & 1 \\ \tau^z & \tau^z \end{smallmatrix} + h_2 \begin{smallmatrix}\tau^x \\ 1 \\ \tau^x \end{smallmatrix}\right).
                \label{eqn:mmm}
            \end{align}

            In the perturbative regime, this is
            \begin{align}
                H_\mathrm{c,e} = - V_1 - &\sum\left(J_0 \begin{smallmatrix} 1 & 1 \\ \tau^z & \tau^z \\ 1 & 1 \end{smallmatrix} + h_0 h_1 \begin{smallmatrix}1 \\ \tau^x \\ 1 \end{smallmatrix}\right) \nonumber \\
                - &\sum \left( J_1 \begin{smallmatrix} \tau^z & \tau^z \\ \tau^z & \tau^z \\ \tau^z & \tau^z \end{smallmatrix} - h_0 h_2  \begin{smallmatrix}1 \\ \tau^x \\ \tau^x \end{smallmatrix}\right) \nonumber \\
                - &\sum\left(J_2 \begin{smallmatrix} 1 & 1 \\ 1 & 1 \\ \tau^z & \tau^z \end{smallmatrix} + h_1 h_2 \begin{smallmatrix}1 \\ 1 \\ \tau^x \end{smallmatrix}\right) + \cdots.
            \end{align}
            This bears some similarity to the Ashkin-Teller model if we disregard the top layer \cite{Ashkin1943,Solyom1981,Bridgeman2015}. We have not pursued this comparison.

        \subsubsection{Condense \(\epsilon^0 \epsilon^1\) in \(N=2\) layers}

            Note that while \(\epsilon^i\) is a fermion, \(\epsilon^0 \epsilon^1\) is a boson.

            We choose \(a^0 = \epsilon^0 \epsilon^1\), \(a^1 = m^0 e^1\), \(b^0 = m^0\), \(b^1 = m^0 m^1\), giving an edge Hamiltonian
            \begin{align}
                H_\mathrm{c,e} = - V_1 - &\sum\left(J_0 \begin{smallmatrix} \tau^y & \tau^z \\ \tau^z & \tau^y \end{smallmatrix} + h_0 \begin{smallmatrix}\tau^x \\ 1 \end{smallmatrix}\right) \nonumber \\
                - & \sum\left(J_1 \begin{smallmatrix} \tau^x & 1 \\ \tau^z & \tau^z \end{smallmatrix} + h_1 \begin{smallmatrix}\tau^x \\ \tau^x \end{smallmatrix}\right).
            \end{align}
            In the perturbative regime \(J_l,h_l \ll 1\) this is again a single copy of the transverse field Ising model.

        \subsubsection{Condense \(\epsilon^0 \epsilon^1\) and \(\epsilon^1 \epsilon^2\) in \(N=3\) layers}

            We choose \(a^0 = \epsilon^0 \epsilon^1\), \(a^1 = \epsilon^1 \epsilon^2\), \(a^2 = m^0 m^1 m^2\) and \(b^0 = e^1 e^2\), \(b^1 = e^0 e^1\), \(b^2 = e^0 e^1 e^2\).

            We have been keeping this discussion at the level of anyons as much as possible. However, that is no longer sufficient for this case. The issue is that while \(\epsilon^0 \epsilon^1\) and \(\epsilon^1 \epsilon^2\) are mutual bosons, \(\epsilon^1\) is a fermion. Thus not all of the movement operators for \(\epsilon^1\) commute. Our assignment of the degrees of freedom in terms of a set of commuting operators requires us to make a choice of which movement operators we will use such that everything commutes. This is most easily achieved by choosing a different definition of the \emph{sites} that support the \(\epsilon^1\) fermion in each of \(a^0\) and \(a^1\).

            First, we will use \(p\) only to refer to white plaquettes in the colouring of \autoref{fig:wen_lattice}, and use \(\tilde{p}\) to refer to the shaded plaquette to the east of \(p\). To distinguish the two choices of sites, we will denote an \(\epsilon\) anyon that resides on faces \(p\) and \(\tilde{p}\) by \(\epsilon\), and one on faces \(p\) and \(\widetilde{p-1}\) by \(\bar{\epsilon}\). The movement operators for \(\epsilon\) commute with all movement operators for \(\bar{\epsilon}\).

            We choose \(a^0 = \bar{\epsilon}^0 \bar{\epsilon}^1\), and the rest as previously stated. This gives a Hamiltonian
            \begin{align}
                H_\mathrm{c,e} = - V_2 - &\sum\left(J_0\begin{smallmatrix} 1 & \tau^x \\ \tau^z & \tau^z \\ \tau^z & \tau^z \end{smallmatrix} + h_0 \begin{smallmatrix}\tau^x \\ 1 \\ \tau^x \end{smallmatrix}\right)  \nonumber\\
                - &\sum\left(J_1\begin{smallmatrix} \tau^z & \tau^y \\ \tau^y & \tau^z \\ \tau^z & \tau^z \end{smallmatrix} + h_1 \begin{smallmatrix}\tau^x \\ \tau^x \\ \tau^x \end{smallmatrix}\right) \nonumber \\
                - &\sum\left(J_2\begin{smallmatrix} \tau^z & \tau^z \\ \tau^x & 1 \\ \tau^z & \tau^z \end{smallmatrix} + h_2 \begin{smallmatrix}1 \\ \tau^x \\ \tau^x \end{smallmatrix}\right).
            \end{align}

        \subsubsection{Condense \(\epsilon^{i-1} m^i \epsilon^{i+1}\) in \(N\) layers}

            Reference~\cite[Section IV]{Wang2013} describes a three-dimensional symmetry protected topological phase (SPT) through a condensation of \(\epsilon^{i-1} m^i \epsilon^{i+1}\) (\(i \in \{1,\ldots, N-2\}\)) in \(N \geq 5\) layers of the toric code. We will show how to obtain the corresponding edge model for \(N \not\equiv 0 \mod 3\).

            We choose \(a^i = \epsilon^{i-1} m^i \bar{\epsilon}^{i+1}\) (note the bar on \(\epsilon^{i+1}\)), interpreting \(i \mod N\). The form of \(b\) depends on whether \(N\equiv 1\) or \(2 \mod 3\). We put
            \begin{equation}
                b^i = \begin{smallmatrix} \vdots \\ 1 \\ e \\ e \\ 1 \\ e^i \\ 1 \\ e \\ e \\ 1 \\ \vdots \end{smallmatrix},
                \quad \text{or,} \quad
                b^i = \begin{smallmatrix} \vdots \\ e \\ e \\ 1 \\ e \\ e^i \\ e \\ 1 \\ e \\ e \\ \vdots \end{smallmatrix},
            \end{equation}
            with the first being for \(N \equiv 2 \mod 3\). The form of this is repeating blocks of \(ee1\), with the specific structure near layer \(k\) depending on \(N \mod 3\).

            Only the Hamiltonian for the \(N\equiv 2 \mod 3\) case will be reproduced here for the sake of brevity. It is
            \begin{equation}
                H_\mathrm{c,e} = - k\sum_{l=1}^{n-2} \sum_q \tau^{z,l}_q - \sum_{l} \sum_q
                \left( J_l \begin{smallmatrix}  \vdots & \vdots \\ 1 & 1 \\ \tau^z & \tau^z \\ \tau^z & \tau^z \\ 1 & \tau^x \\ \tau^{z,l}_q & \tau^{z,l}_{q+1} \\ \tau^x & 1 \\ \tau^z & \tau^z \\ \tau^z & \tau^z \\ 1 & 1 \\ \vdots & \vdots  \end{smallmatrix} 
                +h_l \begin{smallmatrix} \tau^x \\ \tau^{x,l}_q \\ \tau^x \end{smallmatrix}\right).
            \end{equation}
            Our choice of anyons results in the condensation occurring in the middle layers, reflecting the physical situation. The position indices have been omitted on many of the \(\tau\) operators, as we did previously. The pattern of \(\tau^z \tau^z\) terms in the large operator above should match the pattern of \(e\) anyons in \(b^k\), while the \(\tau^x\) part should always be as it appears here. This is also true for the \(N \equiv 1 \mod 3\) case.

            It is not easy to see what difference this has to our previous examples, but in the small \(J/k\), small \(h/k\) regime this should be the edge theory of two copies of a time-reversal symmetry respecting version of the three-fermion model \cite[Section IV]{Wang2013}. Understanding how this theory differs from two copies of the transverse field Ising model would likely be enlightening.

\section{Chiral example: Kitaev spin liquids}
    \label{sec:KSL}

    The characterisation of anyon condensation in \autoref{sec:general} is particularly useful when the bulk phase is chiral. In this case, the edge modes are necessarily gapless, with low energy excitations being described by a CFT in the simplest cases~\cite{Ginsparg1988,Francesco1997,Tong2016}. However, for a given anyon theory in the bulk there is a countable infinity of candidate CFTs that could appear on the edge~\cite{Bais2009,Kitaev2011,Lan2016,Burnell2018}. Which of these CFTs is the correct one is fixed by the knowledge of the chiral central charge, \(c_-\).

    The chiral central charge does not change when a condensation phase transition occurs (\autoref{subsec:central_charge}). This observation fixes the topological phase that is achieved after condensation. Furthermore, the resulting edge CFT can be characterised in terms of the pre-condensation CFT and the introduction of additional local operators (in technical language, extending the chiral algebra~\cite{Bais2009,Burnell2018}).

    In this section, we will illustrate these points through a detailed discussion of anyon condensation in the Kitaev spin liquid (KSL) phases~\cite{Kitaev2006}. These are a relatively simple class of chiral (in general) topological phases which have been widely studied in the literature~\cite{Yao2007,Yao2010,Hermanns2018,Urban2018bilayerKit,Takagi2019}. In \autoref{sec:KSL_back}, we will review their classification in terms of a Chern number \(\nu\) and their basic properties, with a particular focus on \(\nu=1\). Then we will proceed to deduce the edge physics of higher \(\nu\) KSLs by treating them as the result of a condensation in first two (\autoref{sec:KSL_cond}) then several (\autoref{sec:KSL_multi}) layers of \(\nu=1\). Lastly, we discuss an alternative condensation phase transition that can occur when \(\nu \equiv 0 \mod 16\) in \autoref{sec:E8}. The result is a nontrivial topological phase which, nonetheless, has no anyons, called the \(E_8\) state~\cite{Bais2009,Kitaev2011,Lan2016,Burnell2018}.

    \subsection{Kitaev spin liquids}
        \label{sec:KSL_back}

        The Kitaev spin liquids (KSLs) are a class of two dimensional bosonic topological phases of matter. They have a relatively simple structure while still allowing for the presence of non-Abelian anyons and chiral edge modes~\cite{Kitaev2006,Hermanns2018,Takagi2019}.

        The KSLs are characterised by an integer topological invariant usually denoted \(\nu\). This invariant is related to the chiral central charge through
        \begin{equation}
            c_- = \frac{\nu}{2}.
            \label{eqn:KSL_nu_c}
        \end{equation}
        As such, \(\nu\) may be interpreted as counting the number of \emph{fermionic} right-movers at the edge of the system---which is twice the number of bosonic degrees of freedom~\cite{Ginsparg1988}.

        The anyon theory describing the KSLs is determined by \(\nu\) modulo \(16\) (an illustration of the fact that distinct topological phases may have the same anyon theory). The description of the anyons is most easily discussed by further breaking up the models into finer classes.

        \(\nu \equiv 0 \mod 4\).---The even \(\nu\) cases are very similar, and, in fact, only differ in their fusion rules. When \(\nu\) is divisible by \(4\) the anyons are usually labeled
        \begin{equation}
            1 \; \text{(vacuum)}, \quad
            \epsilon \; \text{(fermion)}, \quad
            e, \, m \; \text{(vortices)},
        \end{equation}
        which have the same fusion rules as the toric code
        \begin{multline}
            \epsilon \times \epsilon = e \times e = m \times m = 1, \quad
            e \times \epsilon = m, \\
            m \times \epsilon = e, \quad \text{and}\quad
            e \times m = \epsilon.
        \end{multline}
        Indeed, \(\nu=0\) is the toric code phase.

        The topological spins are determined by \(\nu\):
        \begin{equation}
            \theta_1 = 1, \quad
            \theta_\epsilon = -1, \quad
            \theta_e = \theta_{m} = e^{2\pi i \tfrac{\nu}{16}}.
        \end{equation}
        The remaining nontrivial monodromies not implied by the spins of the anyons are
        \begin{equation}
            R_{m}^{e \epsilon} R_{m}^{\epsilon e} = -1, \quad
            R_{\epsilon}^{e m} R_{\epsilon}^{m e} = -e^{2\pi i \tfrac{2\nu}{16}},
        \end{equation}
        and a similar vortex-fermion braiding for \(m\) and \(\epsilon\). In words, fermions acquire a \((-1)\) phase when wrapped around a vortex, and vortices acquire a phase that depends on \(\nu\) when one is wrapped around another~\footnote{Kitaev~\cite{Kitaev2006} makes a distinction between \(\nu \equiv 0 \mod 8\) and \(\nu \equiv 4\mod 8\), but in the combinations \(R_{m}^{e \epsilon} R_{m}^{\epsilon e}\) and \(R_{\epsilon}^{e m} R_{\epsilon}^{m e}\) the distinction is not required. More complicated braids may need to distinguish these cases.}.

        \(\nu \equiv 2 \mod 4\).---As noted above, the case of \(\nu \equiv 2 \mod 4\) differs from \(\nu \equiv 0 \mod 4\) only in the fusion rules. To distinguish this behaviour, an alternative labeling of the vortices is usually used:
        \begin{equation}
            1 \; \text{(vacuum)}, \quad
            \epsilon \; \text{(fermion)}, \quad
            a, \, \bar{a} \; \text{(vortices)},
        \end{equation}
        which have the nontrivial fusion rules
        \begin{multline}
            \epsilon \times \epsilon = 1, \quad
            a \times \epsilon = \bar{a}, \quad
            \bar{a} \times \epsilon = a, \\
            a \times a = \bar{a} \times \bar{a} = \epsilon, \quad \text{and}\quad
            a \times \bar{a} = 1.
        \end{multline}

        The topological spins are again determined by \(\nu\):
        \begin{equation}
            \theta_1 = 1, \quad
            \theta_\epsilon = -1, \quad
            \theta_a = \theta_{\bar{a}} = e^{2\pi i \tfrac{\nu}{16}}.
        \end{equation}
        The remaining nontrivial braidings are
        \begin{equation}
            R_{\bar{a}}^{a \epsilon} R_{\bar{a}}^{\epsilon a} = -1, \quad
            R_{1}^{a \bar{a}} R_{1}^{\bar{a} a} = e^{-2\pi i \tfrac{2\nu}{16}},
        \end{equation}
        and a similar vortex-fermion braiding for \(\bar{a}\) and \(\epsilon\).

        \(\nu \equiv 1 \mod 2\).---When \(\nu\) is odd the anyon model is non-Abelian. The anyons are labeled
        \begin{equation}
            1 \; \text{(vacuum)}, \quad
            \epsilon \; \text{(fermion)}, \quad
            \sigma \; \text{(vortex)},
        \end{equation}
        which have the nontrivial fusion rules
        \begin{equation}
                \epsilon \times \epsilon = 1, \quad
                \sigma \times \epsilon = \sigma, \quad \text{and}\quad
                \sigma \times \sigma = 1 + \epsilon.
        \end{equation}
        Recall that the ``\(+\)'' here should be thought of as a direct sum. Two \(\sigma\) anyons fuse to the vacuum or a fermion depending on the state of the system globally.

        The topological spins of each species are
        \begin{equation}
            \theta_1 = 1, \quad
            \theta_\epsilon = -1, \quad \text{and}\quad
            \theta_\sigma = e^{2\pi i \tfrac{\nu}{16}}.
        \end{equation}
        The nontrivial braidings relations are those for \(\epsilon\) and \(\sigma\), which acquire a minus sign on braiding,
        \begin{equation}
            R_{\epsilon}^{\epsilon \sigma} R_{\epsilon}^{\sigma \epsilon} = -1,
        \end{equation}
        and that for exchanging vortices, which gives a phase which depends on whether they fuse to \(1\) or \(\epsilon\), and on \(\nu\):
        \begin{equation}
            R_{1}^{\sigma\sigma} = \chi e^{-2\pi i \tfrac{\nu}{16}}, \quad
            R_{\epsilon}^{\sigma\sigma} =  \chi e^{2\pi i \tfrac{3 \nu}{16}}.
        \end{equation}
        Here \(\chi = e^{2\pi i \tfrac{\nu^2-1}{16}}\) is given by \(\chi=1\) for \(\nu \equiv 1,7 \mod 8\) and \(\chi=-1\) for \(\nu \equiv 3,5 \mod 8\).

        Exactly soluble models are known that achieve several different values of \(\nu\)~\cite{Kitaev2006,Yao2007,Yang2007,Kells2011}. Indeed, the achiral phase \(\nu=0\) is actually the toric code phase. However, the exact solubility of these models is meant in a more limited sense than occurs in the toric code. Usually, these models have immobile vortex anyons, and may be mapped to a model of free fermions in any specific background of frozen vortices. This allows one, in principle, to find all the energies and eigenstates, but the vortex movement operators do not have a simple closed form. This should be expected---chiral models have nonzero correlation lengths, and so not all the movement operators can be simple strings of finite range operators.

        As a consequence, our characterisation of the KSL edge physics must be presented with a greater level of abstraction than for the toric code. We will not to find an explicit model for the edge that demonstrates the features we predict, and only derive a more schematic description.

        \subsubsection{The \(\nu=1\) Kitaev spin liquid}

            The exactly soluble models for the KSLs will still be useful for us, as they allow a straightforward identification of the edge physics for \(\nu=1\). We can then use our techniques to build up the description of higher \(\nu\) edges.

            Specifically, Kitaev's honeycomb model~\cite{Kitaev2006,Hermanns2018,Takagi2019} can, in the vortex free sector, be mapped to a model of noninteracting Majorana fermions with a Chern number of \(\nu \in \{0,\pm 1\}\). Chern insulators have \(\nu\) chiral modes at their edge. As the model is one of Majorana fermions, the edge mode is also Majorana. Thus the CFT describing the edge of the \(\nu=1\) KSL is that with a single right-moving Majorana fermion.

            It is useful to see how the bulk anyon content is reflected in the edge CFT explicitly for the \(\nu=1\) KSL, as similar structures occur in later examples.

            The free Majorana CFT has chiral central charge \(c_- = 1/2\)~\cite{Ginsparg1988}, as expected from Eq.~\eqref{eqn:KSL_nu_c}. Being a \emph{minimal model}, the free fermion CFT has finitely many (four) Virasoro primaries. They are
            \begin{equation}
                1 \; \text{(vacuum)}, \quad
                \psi \; \text{(fermion)}, \quad
                \sigma,\, \mu \; \text{(twists)},
            \end{equation}
            with operator product expansions (OPEs) (\autoref{sec:background})
            \begin{multline}
                [\psi][\psi] = [\sigma][\sigma] = [\mu][\mu] = [1], \quad
                [\psi][\sigma] = [\mu], \\
                [\psi][\mu] = [\sigma], \quad\text{and}\quad
                [\sigma][\mu] = [\psi].
            \end{multline}
            The conformal weights of each nontrivial primary are \(h_\psi = \half\) and \(h_{\sigma} = h_{\mu} = \tfrac{1}{16}\)~\cite{Ginsparg1988}. The anti-holomorphic weights are all zero---all these fields are right-moving, not left-moving.

            The fusion rules and conformal weights can be used to deduce braiding relations by matching scaling dimensions. For example, matching scaling dimensions in \([\psi][\sigma] = [\mu]\) gives~\eqref{eqn:OPE}
            \begin{equation}
                \psi(z)\sigma(w) = \frac{1}{(z-w)^{1/2}} \mu(z) + \cdots
            \end{equation}
            from which we can see that moving \(z\) around \(w\) introduces a minus sign due to the branch cut in the square root. Similarly, we have for braiding \(\sigma\) and \(\mu\)
            \begin{equation}
                \sigma(z)\mu(w) = \frac{1}{(z-w)^{-3/8}} \psi(z) + \cdots
            \end{equation}

            The structure of these primary fields is very suggestive of the bulk anyon structure we described for the \(\nu=1\) KSL. More explicitly, we have the following correspondence between bulk anyons and edge primaries, induced by taking a bulk anyon to the edge:
            \begin{equation}
                1 \leftrightarrow [1], \quad
                \epsilon \leftrightarrow [\psi], \quad
                \sigma \leftrightarrow [\sigma] \text{ or } [\mu].
            \end{equation}
            Whether \(\sigma\) (the vortex anyon) produces \([\sigma]\) (the twist field and its descendants) or \([\mu]\) when brought to the edge depends on the global state of the bulk system. Namely, what pairs of vortices fuse to \(\epsilon\) or \(1\).

            One can see that this correspondence is legitimate by checking the fusion and braiding rules. Some fusion rules are immediate, \(\epsilon \times \epsilon = 1 \leftrightarrow [\psi][\psi] = [1]\), but some require us to understand when \(\sigma\) produces \([\sigma]\) or \([\mu]\).

            If we have a pair of anyons \(\sigma\) in the bulk fusing to \(1\), then on the edge they should correspond to \([\sigma]\) and \([\sigma]\) (or \([\mu]\) and \([\mu]\), which is equivalent through the creation of a pair of fermions from the vacuum). In this fermion parity even sector, all the fusion and braiding rules are obeyed:
            \begin{multline}
                \sigma \times \sigma = 1 \quad\text{and}\quad (R_1^{\sigma \sigma})^2 = e^{-2\pi i \tfrac{2}{16}} \\
                \longleftrightarrow \quad
                \sigma(z) \sigma(w) = (z-w)^{-1/8} 1 + \cdots
            \end{multline}
            (We focus on the monodromy \((R_1^{\sigma \sigma})^2\), representing a loop of one vortex around another---rather than \(R_1^{\sigma \sigma}\), representing exchange of vortices---as it is simpler.)
            
            The bulk fusion rule \(\epsilon \times \sigma = \sigma\) should be thought of as appending a fermion to \(\sigma\), which can be resolved at the edge if we are decomposing primaries there into parity even and odd sectors. The addition of a fermion \(\epsilon \leftrightarrow [\psi]\) to one of the vortices brings us to the fermion parity odd sector, where the \(\sigma\) anyons now fuse to \(\epsilon\), and correspond at the edge to \([\sigma]\) and \([\mu]\) (or vice versa, again by the creation of a pair of fermions). This follows from the edge product \([\psi][\sigma] = [\mu]\). In this sector, the fusion and braiding rules are once more obeyed:
            \begin{multline}
                \sigma \times \sigma = \epsilon \quad\text{and}\quad (R_\epsilon^{\sigma \sigma})^2 = e^{2\pi i \tfrac{2 \cdot 3}{16}} \\
                \longleftrightarrow \quad
                \sigma(z) \mu(w) = (z-w)^{3/8} \psi(z) + \cdots
            \end{multline}

    \subsection{Condensation in two layers of \(\nu=1\)}
        \label{sec:KSL_cond}

        The edge physics of the \(\nu=1\) KSL is straightforwardly deduced from the exactly soluble honeycomb model~\cite{Kitaev2006}. There are similar exactly soluble models that achieve other values of \(\nu\)~\cite{Yang2007,Kells2011}, but it becomes progressively more difficult to characterise the edge without some additional knowledge of CFT. In this section, we show that the \(\nu=2\) edge CFT may be characterised by applying the methods of \autoref{sec:general} to a condensation transition in a double layer of \(\nu=1\).

        Anyons in a model with multiple layers are composites of anyons in each layer. In a double layer of the \(\nu=1\) KSL, the composite \(\epsilon_1 \epsilon_2\) consisting of fermions on each layer is a boson. That is, its topological spin is unity,
        \begin{equation}
            \theta_{\epsilon_1 \epsilon_2} = \theta_{\epsilon_1} \theta_{\epsilon_2} = (-1)^2 = 1.
        \end{equation}
        Consequently, it is a legitimate target for condensation~\cite{Burnell2018}.

        The phase resulting from the condensation of \(\epsilon_1 \epsilon_2\) is the \(\nu=2\) KSL~\cite{Burnell2018}. We can see this at the level of anyons by carrying through the algebraic transformation of the bulk anyon theory---confinement, identification, and splitting. Anything which does not have trivial statistics with \(\epsilon_1 \epsilon_2\) becomes confined. The vortices on each layer, \(\sigma_1\) and \(\sigma_2\), each gain a factor of \(-1\) upon braiding with \(\epsilon_1 \epsilon_2\), and so are both confined. However, the composite vortex \(\sigma_1 \sigma_2\) remains deconfined. Similarly, the individual fermions, \(\epsilon_1\) and \(\epsilon_2\), remain deconfined.

        All anyons related by fusion of \(\epsilon_1 \epsilon_2\) should be identified, so actually the fermions on each layer become a single species, \(\epsilon_1 \sim \epsilon_2 \sim \epsilon\). Microscopically, we can think of \(\epsilon_1 \epsilon_2\) creation operators as hopping a fermion from one layer to the other, so it makes sense that \(\epsilon_1\) should be able to transform into \(\epsilon_2\).

        Lastly, we come to splitting. The composite vortex \(\sigma_1 \sigma_2\) is not one of the ``irreducible'' anyons of this model. (Recall that this follows from the fact that \(\sigma_1 \sigma_2\) and its antiparticle, which is also \(\sigma_1 \sigma_2\), can fuse to the condensed anyon \(\epsilon_1 \epsilon_2\).) The composite vortex can be decomposed into a direct sum of two Abelian anyons while preserving all fusion and braiding rules. These are the \(\nu=2\) vortices,
        \begin{equation}
            \sigma_1 \sigma_2 = a + \bar{a}.
        \end{equation}
        One may check that this identification conforms with the fusion and braiding rules for both \(\nu=2\) and two layers of \(\nu=1\).

        In \autoref{subsec:nu2_prime}, we derive another way of showing the condensate phase is \(\nu=2\) by examining the primary fields at the edge.

        What is the effective theory describing the edge of the resulting \(\nu=2\) phase? An expert may be aware, or correctly intuit, that it is a free boson CFT. We will be able to deduce this from our construction.

        \subsubsection{Symmetry breaking}

            As always, there is a symmetry associated to the anyons of the bulk which acts on the edge. It is convenient to use the picture of an anyon existing past the edge, the topological charge of which is revealed by the loop operators \(\Lcal^X_a\) (\autoref{fig:punctured_plane}).

            Before any condensation, the edge theory is two decoupled free fermion CFTs, which each have an independent symmetry algebra. The loop \(\Lcal^X_{\epsilon_l}\) reveals whether the layer \(l\) edge theory has periodic (the topological charge of the edge is \(1\) or \(\epsilon\)) or antiperiodic (the topological charge is \(\sigma\)) boundary conditions for the free fermion. The full edge theory is a direct sum of both of these sectors.

            Meanwhile, \(\Lcal^X_{\sigma_l}\) reveals the fermion parity of the edge. In fact, when the topological charge of the edge is \(\sigma\), the eigenvalue of this loop operator even distinguishes between the case where the anyon in \(\Lcal^X_{\sigma_l}\) fuses with the edge to give \(1\) or \(\epsilon\), as these fusion channels have different braiding phases. If we did not already know the edge CFT was a free fermion, and that the twist operators could be split up into fermion parity even, \([\sigma]\), and odd, \([\mu]\), this observation may have let us deduce this fact.

            When \(\epsilon_1 \epsilon_2\) is condensed the fermion parity in each layer is no longer conserved---only the global fermion parity remains conserved. Explicitly, we may now add \(\Lcal^Y_{\epsilon_1 \epsilon_2}(x)\) to the edge Hamiltonian, breaking the \(\Lcal^X_{\sigma_l}\) symmetries. \(\Lcal^Y_{\epsilon_1 \epsilon_2}(x)\) is a pair creation operator for fermions \emph{on different layers}. Fermions may now move between layers, which alters the fermion parity of each of the separate layers.

            From this perspective, the trivial action of \(\Lcal^X_{\epsilon_1 \epsilon_2}\) in the condensate phase also becomes natural. Eigenvalues of \(\Lcal^X_{\epsilon_1 \epsilon_2}\) reveal the difference in the boundary conditions between the fermions in layer 1 and those in layer 2. But fermions in layer 1 can now hop to layer 2, so this distinction no longer makes sense, and the boundary conditions must be the same.

        \subsubsection{Invariance of the central charge}

            The chiral central charge of the free fermion CFT is \(c_- = 1/2\). As such, the central charge for the double-layer model is 
            \begin{equation}
                c_- = 2 \cdot(1/2) = 1.
            \end{equation}\(\)
            Together with the characterisation of the bulk anyons, this observation confirms that the bulk phase after condensation is, indeed, the \(\nu=2\) KSL, as we have claimed.

            While knowledge of the central charge is helpful and necessary, there is not a unique CFT with \(c_- = 1\), so a more complete description of the edge CFT requires more work.

        \subsubsection{Extension of the chiral algebra}
            \label{subsec:nu2_prime}

            The free fermion CFT has only four Virasoro primaries, and only the vacuum \([1]\) is local. The chiral algebra, in our intuitive description as the algebra of local operators, is trivial, containing only the descendants of the vacuum. After condensation, the composite field \([\psi_1 \psi_2]\) and its descendants are included in the chiral algebra. This lets us make a more complete characterization of the primaries appearing in the new edge CFT.

            It is convenient to bosonise the edge CFT of the two-layer model~\cite{Kitaev2011}. Nominally, the edge is described by a CFT of two free Majorana fermions \(\psi_1\) and \(\psi_2\). However, it is possible to express these in terms of a boson \(\phi\) as~\cite{Ginsparg1988}
            \begin{equation}
                \psi_{\pm} = \frac{i}{\sqrt{2}}(\psi_1 \pm i \psi_2) = e^{\pm i \phi}.
            \end{equation}

            Of special interest in this CFT are the (Virasoro) primaries known as the vertex operators,
            \begin{equation}
                V_\alpha(z) = e^{i \alpha \phi(z)}.
            \end{equation}
            With free boundary conditions all \(\alpha \in \C\) give inequivalent primaries. With periodic or anti-periodic boundary conditions only \(\alpha \in \Z \cup (\Z+\half)\) are acceptable. There are now infinitely many Virasoro primaries.

            The OPE and braiding of vertex operators is given by
            \begin{equation}
                V_\alpha(z) V_\beta(w) = (z-w)^{\alpha \beta} V_{\alpha + \beta}(z)(1 + \cdots).
                \label{eqn:vertex}
            \end{equation}

            While all the vertex operators \(V_\alpha\) are Virasoro primaries, we will find they fall into finitely many \emph{chiral} equivalence classes. That is, many are related by operators that can be applied entirely locally. As anyons cannot have their species changed locally, bulk anyons correspond to an entire family of Virasoro primaries that belong to the same chiral equivalence class.

            The chiral algebra is generated by the fields corresponding to \(\Lcal^Y_{\epsilon_1 \epsilon_2}(x)\). In the CFT, this is \([\psi_1 \psi_2]\). We observe that
            \begin{subequations}
            \begin{align}
                [\psi_+][\psi_+] &= -\frac{1}{2}([\psi_1][\psi_1] - [\psi_2][\psi_2] + 2i[\psi_1][\psi_2]) \\
                &= -i[\psi_1][\psi_2],
            \end{align}
            \end{subequations}
            so that \(\psi_+^2 = V_2\) should be in the chiral algebra. Similarly, \(V_{-2}\) is in the chiral algebra. Then the values of \(\alpha\) corresponding to vertex operators in the chiral algebra are
            \begin{equation}
                \alpha \in 2 \Z.
            \end{equation}
            From Eq.~\eqref{eqn:vertex}, these are precisely the vertex operators that are self-bosons.

            Then chiral equivalence classes correspond to cosets of the form \(\alpha_0 + 2\Z\). There are four such cosets for \(\alpha \in \Z \cup (\Z+\half)\). They have representatives
            \begin{equation}
                \alpha_0 = 0, \quad
                \alpha_0 = 1, \quad
                \alpha_0 = \half, \quad\text{and}\quad
                \alpha_0 = -\half.
            \end{equation}
            The vertex operators corresponding to these \(\alpha_0\) are representatives of each of the chiral equivalence classes.

            We can also see how to construct these from the fermion operators in each layer. We have already discussed \(\alpha_0 = 0\)---the chiral algebra itself. These are fermion hops. Equally simple is \(\alpha_0 = 1\). These are the fermion operators, as can be seen from
            \begin{equation}
                \psi_1 \propto V_{1} + V_{-1},
            \end{equation}
            which expresses \(\psi_1\) in terms of the vertex operators \(\psi_\pm = V_{\pm 1}\), which both correspond to an odd \(\alpha \in 1 + 2 \Z\).

            The remaining \(\alpha_0 = \pm 1/2\) correspond to the twist fields, and their descendants. The fermion parity definite twist fields that survived the bulk condensation are (up to fusion with \([\psi_1\psi_2]\))
            \begin{equation}
                [\sigma_{1} \sigma_{2}] \quad\text{and}\quad [\sigma_{1}\mu_{2}].
            \end{equation}
            However, neither of these are candidates for \(\alpha_0 = \pm 1/2\) because, for instance, they both square to 1, not \(\psi\). Rather, we should use the linear combinations
            \begin{equation}
                [\sigma_{\pm}] = \frac{i}{\sqrt{2}}([\sigma_{1} \sigma_{2}] \pm i[\sigma_{1}\mu_{2}]).
                \label{eqn:sigmapm}
            \end{equation}
            That these satisfy all the fusion and braiding rules for an operator in the equivalence class of \(V_{\pm1/2}\). Namely, they both square to a fermion, annihilate each other, have the right spin, and so on.

            This accounts for all the fields arising from bringing an anyon to the edge of the system. We can confirm that the edge CFT we have deduced---that of a free boson, with chiral equivalence classes characterised in terms of vertex operators---is compatible with the bulk anyon theory by checking fusion and braiding relations. (This CFT has the correct value \(c_- = 1\)~\cite{Ginsparg1988}.)

            We state the correspondence between \(\nu=2\) anyons and chiral classes in terms of \(\alpha_0\).
            \begin{multline}
                1 \leftrightarrow \alpha_0 = 0, \quad
                \epsilon \leftrightarrow \alpha_0 = 1, \\
                a \leftrightarrow \alpha_0 = \half, \quad\text{and}\quad
                \bar{a} \leftrightarrow \alpha_0 = -\half.
                \label{eqn:nu2_correspond}
            \end{multline}
            Which vortex gets assigned to \(\pm 1/2\) is arbitrary.

            We discuss each correspondence in Eq.~\eqref{eqn:nu2_correspond}, beginning with \(1 \leftrightarrow 0\)---the vacuum corresponds with the chiral algebra itself. This is definitional: the chiral algebra contains all local operators, while the vacuum anyon also captures all purely local operations.

            The next correspondence, \(\epsilon \leftrightarrow 1\), is also straightforward. It is inherited from \(\epsilon_l \leftrightarrow [\psi_l]\). Recall \(V_{1}\) is a linear combination of \(\psi_1\) and \(\psi_2\), which are identified by the chiral algebra.

            The cases of \(a \leftrightarrow 1/2\) and \(\bar{a} \leftrightarrow -1/2\) are more complicated, but again we have done most of the work. Both \(a\) and \(\bar{a}\) came from \(\sigma_{1}\sigma_{2}\), which corresponds to either \([\sigma_{1} \sigma_{2}]\) or \([\sigma_{1}\mu_{2}]\) at the edge, depending on fermion parity. We saw that the correct combination of these fields to ensure membership of just one chiral equivalence class are \([\sigma_\pm]\) from Eq.~\eqref{eqn:sigmapm}. These are then what \(a\) and \(\bar{a}\) should correspond to. From an alternative perspective, that we can split the edge chiral classes into two distinct cosets \([\sigma_\pm]\) hints that the bulk anyon \(\sigma_1 \sigma_2\) splits into two different anyons.

            The fusion and braiding relations are also satisfied by Eq.~\eqref{eqn:nu2_correspond}. Indeed, from \eqref{eqn:vertex}, we see that fusion is just addition for \(\alpha\). Taking the quotient by the chiral algebra is regarding the result modulo 2. Then we can readily see that
            \begin{subequations}
            \begin{align}
                \epsilon \times \epsilon = a \times \bar{a} = 1 &\leftrightarrow 1 + 1 = \half - \half = 0 \mod 2, \\
                \epsilon \times a = \bar{a} &\leftrightarrow 1 + \half = -\half \mod 2, \\
                \epsilon \times \bar{a} = a &\leftrightarrow 1 - \half = \half \mod 2, \quad\text{and}\\
                a \times a = \bar{a}\times\bar{a} = \epsilon &\leftrightarrow \half +\half = -\half -\half = 1 \mod 2.
            \end{align}
            \end{subequations}

            Obtaining braiding statistics is also immediate, involving multiplication of \(\alpha\) rather than addition. As an example,
            \begin{multline}
                a \times \bar{a} = 1 \quad\text{and}\quad R_1^{a\bar{a}}R_1^{\bar{a} a} = e^{-2\pi i \tfrac{4}{16}} \\
                \longleftrightarrow
                V_{1/2}(z)V_{-1/2}(w) = (z-w)^{-1/4} V_0(z).
            \end{multline}

            The prescription dictated by the two edge model successfully deduces the edge CFT for \(\nu=2\) by extending the chiral algebra in two layers of \(\nu=1\).

    \subsection{Condensation in many layers of Kitaev spin liquids}
        \label{sec:KSL_multi}

        The analysis of \autoref{sec:KSL_cond} is not limited to \(\nu=2\). We can use the same construction in any number of layers to deduce the edge CFT for all Kitaev spin liquids. In this section, we repeat the analysis of \autoref{sec:KSL_cond} (in lesser detail) for positive \(\nu\). The results for negative \(\nu\) are obtained through the action of time reversal.

        The anyons for \(\nu=k>0\) can be obtained by condensing the two-fermion composites \(\epsilon_{l}\epsilon_{l+1}\) in \(k\) layers of \(\nu=1\). Furthermore, \(c_-\) for the resulting topological phase is, by the invariance of the chiral central charge under condensation, \(c_- = \nu/2\). These two facts imply that a KSL of any given positive \(\nu\) may be obtained by condensing pairs of fermions in a stack of \(\nu=1\) KSLs. Using this characterisation of KSLs, it is possible to deduce the edge theory of a KSL for any value of \(\nu\) using the two edge model.

        Namely, condensation of \(\epsilon_{l}\epsilon_{l+1}\) breaks the fermion parity symmetry of each layer. The surviving symmetry group is the global fermion parity across all layers. As occurred for \(\nu=2\), vortices thread all layers after condensation. Thus the periodic or antiperiodic boundary conditions of the edge theory are shared between all layers.

        For \(\nu \geq 2\) there are several candidate CFTs for the edge theory with the correct value of \(c_-\) and symmetry properties. Charaterising the CFT primaries is necessary to distinguish which CFT is correct. Thus the remainder of this section focuses on the description of the primary fields.

        \subsubsection{Even number of layers}
            \label{subsec:KSL_even}

            The KSLs with odd \(\nu\) have non-Abelian anyons~\cite{Kitaev2006}, which complicates their description. By expressing the \(\nu=2n\) KSLs as a condensation of \(n\) layers of \(\nu=2\), we avoid this complication for as long as possible.

            Recall from \autoref{subsec:nu2_prime} that the edge CFT for \(\nu=2\) is a free boson, and may be described in terms of vertex operator primaries. Composites of these vertex operators between the \(n\) layers can be written
            \begin{equation}
                V_{\vec{\alpha}}(z) = e^{i\vec{\alpha} \cdot \vec{\phi}(z)},
            \end{equation}
            where now, with periodic or antiperiodic boundary conditions on each layer, \(\vec{\alpha} \in \left(\Z\cup(\Z+\half)\right)^{n}\), and \(\vec{\phi}(z) = (\phi_1(z), \ldots, \phi_{n}(z))\). The OPE of two such fields takes the form
            \begin{equation}
                V_{\vec{\alpha}}(z)V_{\vec{\beta}}(w) = (z-w)^{\vec{\alpha}\cdot\vec{\beta}}V_{\vec{\alpha}+\vec{\beta}}(z)(1 + \cdots).
                \label{eqn:vertex_multi}
            \end{equation}

            Just as was the case for \(\nu = 2\), some primaries in the uncoupled CFT consisting of \(n\) bosons become gapped when the symmetry protecting them is broken. Anything which is charged under \(\Lcal^X_{\epsilon_l \epsilon_k}\) becomes gapped by \(\Lcal^Y_{\epsilon_l \epsilon_k}(x)\), and disappears from the CFT. This is the same as demanding that the elements of the chiral algebra braid trivially with all primaries. Only \(V_{\vec{\alpha}}(x)\) such that
            \begin{equation}
                \alpha_l + \alpha_k \in \Z \quad \text{for all }l,\,k,
                \label{eqn:alpha_comp_cond}
            \end{equation}
            are allowed. Here, \(\alpha_l\) and \(\alpha_k\) are components of \(\vec{\alpha}\). The \(\vec{\alpha}\) which satisfy Eq.~\eqref{eqn:alpha_comp_cond} are
            \begin{equation}
                \vec{\alpha} \in \Z^{n} \cup \left( (\half,\ldots, \half) + \Z^{n} \right).
            \end{equation}
            The first partition of this set consists of bosons and fermions. The half-integer part consists of vortices. That all entries must be half-integer if any of them are is a reflection of the fact that vortices must thread all the layers.

            The chiral algebra may also be characterised by a condition on \(\vec{\alpha}\)~\cite{Kitaev2011}. It is generated by composites of fermions on different layers. The vertex operators obtained by taking products of these may be interpreted as hopping any number of fermions between any layers---equivalently, creating an even number of fermions. Thus the chiral algebra consists of vertex operators \(V_{\vec{\alpha}}\) with
            \begin{multline}
                \vec{\alpha} \in \Z_{\mathrm{even}} 
                = \bigg\{\vec{\alpha} = (\alpha_1, \ldots, \alpha_{n}) \in \Z^{n} \;:\\
                 \sum_l \alpha_l \equiv 0 \mod 2 \bigg\}.
            \end{multline}
            Once more, these are precisely the vertex operators which are bosons. Eq.~\eqref{eqn:vertex_multi} shows that the conformal weight of \(V_{\vec{\alpha}}\) is \(h_{\vec{\alpha}} = |\vec{\alpha}|^2/2\). If \(\vec{\alpha} \in \Z_{\mathrm{even}}\), we have
            \begin{equation}
                |\vec{\alpha}|^2 = \sum_{l} \alpha_l \equiv 0 \mod 2,
            \end{equation}
            as \(\alpha_l^2 \equiv \alpha_l \mod 2\) for integer \(\alpha_l\). Then \(h_{\vec{\alpha}} \in \Z\) is an integer, and \(V_{\vec{\alpha}}\) is a boson.

            This is a complete enough description of the edge CFT for our purposes. We can again check the correspondence between anyons and primaries, which is in complete analogy with \(\nu = 2\). We write \(b \leftrightarrow \vec{\alpha}_{0,b}\) if taking the anyon \(b\) to the edge produces a field in the edge CFT which is a sum of vertex operators \(V_{\vec{\alpha}}\) with \(\vec{\alpha} \in \vec{\alpha}_{0,b} + \Z_{\mathrm{even}}\) and their descendants.

            Then the correspondence inherited from \(\nu=2\) according to the construction of \(\nu = 2n\) by \(n\) layers is
            \begin{multline}
                1 \leftrightarrow (0,0,\ldots,0), \quad
                \epsilon \leftrightarrow (1,0,\ldots,0), \\
                a, e \leftrightarrow (\half,\half,\ldots,\half), \quad\text{and}\quad
                \bar{a}, m \leftrightarrow (-\half,\half,\ldots,\half).
            \end{multline}
            Whether the vortices are labeled \(a\), \(\bar{a}\) or \(e\), \(m\) depends on \(\nu\).

            One may also explicitly check that all the fusion and braiding relations match with this assignment. From our construction, these relations must match, but we will also check this for a small number of examples.

            Indeed, for the fusion and braiding of \(a\) and \(\bar{a}\) in any \(\nu = 2 (2k+1)\), we have
            \begin{multline}
                a \times \bar{a} = 1 \quad\text{and}\quad R_1^{a\bar{a}}R_1^{\bar{a} a} = e^{-2\pi i \tfrac{4(2k+1)}{16}}
                \\ \longleftrightarrow
                V_{(1/2,\ldots,1/2)}(z)V_{(-1/2,\ldots,1/2)}(w) \\
                = (z-w)^{(2k-1)/4} V_{(0,1,\ldots,1)}(z),
            \end{multline}
            where \((0,1,\ldots,1) \in \Z_{\mathrm{even}}\), as \(2k+1\) is odd, and taking \(z\) around \(w\) produces as phase \(e^{2\pi i \tfrac{(2k-1)}{4}} = e^{-2\pi i \tfrac{4(2k+1)}{16}} \).

            For an example with \(\nu = 4 k\), we have
            \begin{multline}
                e \times m = \epsilon \quad\text{and}\quad R_\epsilon^{em}R_\epsilon^{me} = -e^{2\pi i \tfrac{8k}{16}}
                \\ \longleftrightarrow
                V_{(1/2,\ldots,1/2)}(z)V_{(-1/2,\ldots,1/2)}(w) \\
                = (z-w)^{2(k-1)/4} V_{(0,1,\ldots,1)}(z),
            \end{multline}
            where \((0,1,\ldots,1) \in (1,0,\ldots,0) + \Z_{\mathrm{even}}\), as \(2k\) is even, and taking \(z\) around \(w\) produces a phase \(e^{2\pi i \tfrac{2(k-1)}{4}} = -e^{2\pi i \tfrac{8k}{16}} \).

        \subsubsection{Odd number of layers}
            \label{subsec:KSL_odd}

            To fill in the odd values of \(\nu = 2n +1\), we consider condensing \(\epsilon_1\epsilon_2\) in a stack of \(\nu = 1\) and \(\nu=2n\).

            The uncoupled primaries are composites of free fermion primaries and vertex operators in the \(\nu=2n\) edge,
            \begin{multline}
                1 \cdot V_{\vec{\alpha}}(z), \quad
                \psi(z) V_{\vec{\alpha}}(z), \quad
                \sigma(z) V_{\vec{\alpha}}(z), \\
                \text{and}\quad
                \mu(z) V_{\vec{\alpha}}(z),
            \end{multline}
            where \(\vec{\alpha} \in \Z^{n} \cup \left( (\half,\ldots, \half) + \Z^{n} \right)\).

            After condensing \(\epsilon_1 \epsilon_2 \leftrightarrow [\psi V_{(1,0,\ldots,0)}]\) some of these primaries may be gapped by \(\Lcal^Y_{\epsilon_1 \epsilon_2}(x)\). The remaining primaries are
            \begin{subequations}
            \begin{align}
                1 \cdot V_{\vec{\alpha}}(z), \quad &\vec{\alpha} \in \Z^n, \\
                \psi(z) \cdot V_{\vec{\alpha}}(z), \quad &\vec{\alpha} \in \Z^n, \\
                \sigma(z) \cdot V_{\vec{\alpha}}(z), \quad &\vec{\alpha} \in (\half,\ldots, \half) + \Z^{n},  \\
                \mu(z) \cdot V_{\vec{\alpha}}(z), \quad &\vec{\alpha} \in (\half,\ldots, \half) + \Z^{n}.
            \end{align}
            \end{subequations}
            As we saw in the even case, vortices must thread all the layers.

            The new chiral algebra is found by appending \([\psi V_{(1,0,\ldots,0)}]\) to the chiral algebra for \(\nu=2n\). Also as in the even case, the chiral algebra consists of all operators with an even number of fermions. That is, of all the bosons:
            \begin{subequations}
            \begin{align}
                1 \cdot V_{\vec{\alpha}}(z), \quad &\vec{\alpha} \in \Z_{\mathrm{even}}, \\
                \psi(z) \cdot V_{\vec{\alpha}}(z), \quad &\vec{\alpha} \in \Z_{\mathrm{odd}},
            \end{align}
            \end{subequations}
            where \(\Z_{\mathrm{odd}} = (1,0,\ldots,0) + \Z_{\mathrm{even}}\).

            There are four chiral equivalence classes, with representatives
            \begin{multline}
                1 \cdot V_{\vec{0}}(z), \quad
                \psi(z) \cdot V_{\vec{0}}(z), \quad
                \sigma(z) \cdot V_{\vec{\alpha}_{0,\sigma}}(z), \\
                \text{and} \quad \mu(z) \cdot V_{\vec{\alpha}_{0,\sigma}}(z), \quad
            \end{multline}
            where \(\vec{\alpha}_{0,\sigma} = (\half, \ldots, \half)\). The vacuum anyon and the fermion are clearly identified with the first two fields here. The bulk vortex may correspond to one of two twist fields at the edge, depending on fermion parity (as in the \(\nu=1\) case).

            Let us check a few of the fusion and braiding relations. We choose the vortex braiding relations in each of its fusion channels, as these are the least immediate. For \(n = 2k\) we have
            \begin{multline}
                \sigma \times \sigma = 1 \quad\text{and}\quad (R_1^{\sigma \sigma})^2 = e^{-2\pi i \tfrac{2(2n+1)}{16}}
                \\ \longleftrightarrow
                \sigma(z) V_{\vec{\alpha}_{0,\sigma}}(z) \sigma(w) V_{\vec{\alpha}_{0,\sigma}}(w) \\
                = (z-w)^{n/4 - 1/8} V_{(1,\ldots,1)}(z),
            \end{multline}
            where \((1,\ldots,1)\in\Z_{\mathrm{even}}\), as \(n\) is even. Wrapping \(z\) around \(w\) correctly produces a phase \(e^{2\pi i [n/4 - 1/8]} = e^{-2\pi i \tfrac{2(2n+1)}{16}}\).

            Meanwhile, still for \(n = 2k\), the other fusion channel gives
            \begin{multline}
                \sigma \times \sigma = \epsilon \quad\text{and}\quad (R_\epsilon^{\sigma \sigma})^2 = e^{2\pi i \tfrac{6(2n+1)}{16}}
                \\ \longleftrightarrow
                \sigma(z) V_{\vec{\alpha}_{0,\sigma}}(z) \mu(w) V_{\vec{\alpha}_{0,\sigma}}(w)\\
                = (z-w)^{n/4 + 3/8} V_{(1,\ldots,1)}(z).
            \end{multline}
            Wrapping \(z\) around \(w\) produces a phase \(e^{2\pi i [n/4 + 3/8]} = e^{2\pi i \tfrac{6(2n+1)}{16}}\).

            Similar calculations may be performed when \(n = 2k+1\) is odd.

    \subsection{Condensing vortices in \(\nu=16\) and the \(E_8\) state}
        \label{sec:E8}

        When \(\nu \equiv 0 \mod 16\), the bulk anyon theory is that of the toric code~\cite{Kitaev2003,Kitaev2006}. Notably, the vortices \(e\) and \(m\) are both bosonic, and themselves are candidates for condensation. The anyon theory produced by condensing \(e\) (or \(m\)) in the toric code is trivial (\autoref{sec:toric}). However, the invariance of the chiral central charge under condensation also tells us that the topological phase produced by condensing \(e\) in the \(\nu=16\) KSL has
        \begin{equation}
            c_- = 8.
        \end{equation}

        The condensate phase has no anyons in its bulk, but nonetheless has nontrivial topological order, with eight chiral bosonic edge modes~\cite{Kitaev2011}. This topological phase is known as the \(E_8\) state. It is a particularly striking example of the fact that topological phases are not determined solely by their bulk anyon content~\cite{Kitaev2011,Lan2016}.

        The \(E_8\) state is named for a relation to the exceptional simple Lie group \(E_8\). Deducing the CFT describing the edge of this phase reveals what this connection is.

        Before condensation, the edge CFT has vertex operators \(V_{\vec{\alpha}}(z)\) where \(\vec{\alpha} \in \Z^{8} \cup \left( (\half,\ldots, \half) + \Z^{8} \right)\), and the chiral algebra consists of the vertex operators with \(\vec{\alpha} \in \Z_{\mathrm{even}}\).

        The anyon \(e\) corresponds to the chiral equivalence class of
        \begin{equation}
            \vec{\alpha}_{0,e} = \left(\half, \half, \half, \half, \half, \half, \half, \half\right).
        \end{equation}
        This is a boson: \(h_{\vec{\alpha}_{0,e}} = |\vec{\alpha}_{0,e}|^2/2 = 1 \in \Z\). As such, it can be appended to the chiral algebra.

        None of the other primaries braid trivially with this vertex operator, and so they all become gapped, reflecting the lack of anyons in the bulk condensate.  We are left with only a single chiral equivalence class, and a chiral algebra
        \begin{equation}
            \vec{\alpha} \in \Z_{\mathrm{even}} \cup \left( (\half, \ldots, \half) + \Z_{\mathrm{even}}\right) = \Gamma_8.
        \end{equation}
        
        This lattice, \(\Gamma_8\), is the root lattice for the exceptional simple Lie group \(E_8\). This is the origin of the bulk phase's name.

\section{Discussion}
    \label{sec:disc}

    The abstract mathematical tools employed in the study of topological phases maintain little contact with the physical system. While this makes them powerful and general~\cite{Kitaev2006,Kitaev2012,Lan2016,Kong2018,Kong2020,Kong2021}, it also restricts the class of questions which they can answer. Our motivating question is outside the scope of any method which assumes a large gap in the bulk: what happens to the edge modes when an anyon condenses in the bulk?

    The two edge model introduced in \autoref{sec:general} provides a platform in which simple thought experiments reveal answers to this question: the edge generically breaks a symmetry previously imposed by the bulk anyons~\cite{Lichtman2021,Chatterjee2023}, the chiral central charge (related to the heat current at the edge) remains invariant, and additional operators at the edge become local~\cite{Bais2009}.

    The toric code~\cite{Kitaev2003} provides a setting where we can see these principles manifest explicitly in a soluble model. The edge model in that case is just the transverse field Ising model~\cite{Ho2015}. Anyon condensation in the bulk of the toric code generically introduces a longitudinal field to the edge model, breaking its \(\Z_2\) symmetry.

    The Kitaev spin liquids~\cite{Kitaev2006} also admit anyon condensations when they are layered. By applying two edge technology to these chiral examples, we are able to deduce the nature of the CFT describing the edge for any value of the Chern index \(\nu\) describing the Kitaev phases. We can also construct the \(E_8\) state, an important example of a topological phase without bulk anyons, and see the \(E_8\) lattice structure of its edge CFT.

    Further work could also include an examination of what effect \emph{gauging} in the bulk---which is in a sense the inverse operation of condensation---has on the edge. The two edge model should be directly applicable to this scenario. From the CFT analogy, one should discover that the CFT undergoes a process known as orbifolding~\cite{Ginsparg1988,Francesco1997}. Having a physical picture for this process, through the two edge model, would also be helpful.

    In order to adapt the two edge method to other phase transitions, beyond gauging and condensation, one must consider gapless excitations at both edges, and how they can couple to each other. This should certainly be possible, but is likely to become complicated in general.

    It may be possible to carry through similar two edge thought experiments to study the edges of symmetry protected topological phases (SPTs)~\cite{Senthil2015}. Such phases also express anomalous edge states, and characterising these through the two edge construction could prove enlightening. In this context, it would also be interesting to generalise the construction to higher dimensions, as there are many interesting SPTs which are not two dimensional.

    A possible route towards such a generalisation of our model is to rephrase the two edge thought experiment in the modern language of higher form symmetries~\cite{Gomes2023,Nussinov2009sufficient,Nussinov2009symmetry,Gaiotto2015}. The symmetry operators associated to loops of anyons, \(\Lcal^X_a\), are examples of higher form symmetries, and reformulating our conclusions in this language could open a route to straightforward generalization of the two edge construction---both to higher dimensions and to SPTs. Indeed, there has been recent work exploring the connection between higher form symmetries and anyon condensation~\cite{Verresen2022higgs,Thorngren2023higgs,Huxford2023gaining}.

    The two edge construction provides a means of assessing the phase structure of specific lattice models. As a specific example, consider the condensation transition from the \(\nu=16\) Kitaev spin liquid to the \(E_8\) state described in \autoref{sec:E8}. Models of the \(\nu=16\) Kitaev spin liquids are relatively easy to construct from layers of the \(\nu=1\) honeycomb model. As the vortex movement operators in such a model are not known exactly, it is not clear exactly how one should deform the model in order to cause a condensation of the composite vortex anyons, as required to produce the \(E_8\) state~\cite{Kitaev2011}. Determining whether a given model actually achieves the \(E_8\) state or a trivial state can be a difficult task. Distinguishing the \(E_8\) phase from the trivial phase in the bulk requires knowledge of the wave function~\cite{Kim2022}, which is not available in models without exact solutions. However, by using the two edge geometry, it may be possible characterise an effective model of the edge, as was done for the toric code in \autoref{sec:toric}, and hence verify that a proposed model does lie in the \(E_8\) phase. Similarly, deriving effective models of the edge in other lattice models may assist in determining their phase diagrams.

    At a broader level, our results demonstrate the utility of studying topological phases through the application of simple thought experiments. The kinds of thought experiments we perform underlie the construction of the mathematical theory of anyons. Returning to these thought experiments can overcome questions that would otherwise be rendered opaque. Such thought experiments are usually more direct and intuitive to a broad audience, and the study of anyons and topological phases would likely be more accessible if such constructions were provided when they are possible.

\section*{Acknowledgements}

    The authors thank J. Alicea, G. Brennen, F. Burnell, C. Chamon, A. Chandran, D. Else, C. Laumann, G. Vidal, S. Vishveshwara, and D. Williamson for helpful comments and discussions. This work was supported by NSF Grant No. DMR-1752759, AFOSR Grant No. FA9550-20-1-0235, the Laboratory for Physical Sciences (D.M.L.), and the Australian Research Council via the Centre of Excellence in Engineered Quantum Systems (EQUS) Project No. CE170100009 (D.M.L. and A.C.D.). The authors acknowledge the traditional owners of the land on which this work was undertaken at the University of Sydney, the Gadigal people of the Eora Nation.

\bibliography{edge_bibliography}

\end{document}